\documentclass[journal=jacsat,manuscript=abstract]{achemso}
\usepackage[version=3]{mhchem} % Formula subscripts using \ce{}
\usepackage{graphicx}
\usepackage{epsfig}
\usepackage{natbib}
\usepackage{bm}
\usepackage{color}
\author{Ioanna A. Gorbunova}
\affiliation{Ioffe Institute, 26 Polytekhnicheskaya, St.Petersburg, 194021 Russia}
\author{Maxim E. Sasin}
\affiliation{Ioffe Institute, 26 Polytekhnicheskaya, St.Petersburg, 194021 Russia}
\author{Jesus Rubayo-Soneira}
\affiliation{Universidad de La Habana, Instituto Superior de Tecnologías y Ciencias Aplicadas, La Habana, Cuba}
\author{Andrey G. Smolin}
\affiliation{Ioffe Institute, 26 Polytekhnicheskaya, St.Petersburg, 194021 Russia}
\author{Oleg S. Vasyutinskii}
\affiliation{Ioffe Institute, 26 Polytekhnicheskaya, St.Petersburg, 194021 Russia}
\email{osv@pms.ioffe.ru}
\phone{+7 (812) 297 2064}
\fax{+7 (812) 297 1017}
\title[]
  {Two-Photon Excited Fluorescence Dynamics in NADH in water-methanol solutions:
  the Role of Conformational States}

\abbreviations{NADH, NAD, NMNH, NMH, FLIM, AD, NA, UV, SI, TCSPC, IRF, NMR, PCM, PP}
\keywords{NADH, polarized fluorescence, decay times, rotational diffusion, two-photon excitation, femtosecond laser pulses}

\begin{document}

%\begin{tocentry}

%Some journals require a graphical entry for the Table of Contents.
%This should be laid out ``print ready'' so that the sizing of the

%Inside the \texttt{tocentry} environment, the font used is Helvetica
%8\,pt, as required by \emph{Journal of the American Chemical

%The surrounding frame is 9\,cm by 3.5\,cm, which is the maximum
%permitted for  \emph{Journal of the American Chemical Society}
%graphical table of content entries. The box will not resize if the
%content is too big: instead it will overflow the edge of the box.

%This box and the associated title will always be printed on a
%separate page at the end of the document.

%\end{tocentry}

\begin{abstract}
 The dynamics of polarized fluorescence in NADH at 460~nm under two-photon excitation at 720~nm by femtosecond laser pulses in water-methanol solutions has been studied experimentally and theoretically as a function of methanol concentration. A number of fluorescence parameters have been determined from experiment by means of the global fit procedure and then compared with the results reported by other authors. A comprehensive analysis of experimental errors was made. The interpretation of the experimental results obtained was supported by \emph{ab initio} calculations of the structure of NADH and NMNH in various solutions. The main results obtained are as follows.  An explanation of the heterogeneity in the measured decay times in NADH and NMNH has been suggested based on the influence of the internal molecular electric field in the nicotinamide ring on non-radiative decay rates. We suggest that different charge distributions in the \emph{cis} and \emph{trans} configurations result in different internal electrostatic field distributions that lead to the decay time heterogeneity.  The experimental data was fitted satisfactory by two exponents with isotropic decay times $\tau_1$ and $\tau_2$ and a single exponent with the rotational diffusion time $\tau_r$, while the inclusion of additional exponential decays did not improve the fit quality within given experimental errors. A slight but noticeable rise of the decay times $\tau_1$ and $\tau_2$  with methanol concentration was observed and treated as a minor effect of a non-radiative relaxation slowing due to the decrease of solution polarity.  The analysis of the rotational diffusion time $\tau_r$  as a function of methanol concentration on the basis of the Stokes-Einstein-Debye equation allowed for determination of relative concentrations of the folded and unfolded NADH conformations in solutions. The analysis of the fluorescence anisotropy parameters and parameter $\Omega$ determined from experiment allowed for determination of the two-photon excitation tensor components and suggested the existence of two excitation channels with comparable intensities. These were the longitudinal excitation channel dominated by the two-photon diagonal tensor component $S_{zz}$ in the direction $Z$ approximately parallel to the long axis of the nicotinamide ring and the mixed excitation channel dominated by the off-diagonal tensor components $|S^2_{xz}+S^2_{yz}|^{1/2}$.

\end{abstract}

\section{Introduction}

Nicotinamide adenine dinucleotide (NAD) is an important natural biological cofactor essential for regulation of redox
reactions and metabolic pathways in living cells that is intensively used for decades as an indicator of the metabolic
states~\cite{Chance1962,Heiden2009,Belenky2007}. As known,~\cite{Pollak2007} hydride-carrying processes occur between
substrate and a nicotinamide ring in the course of redox reactions resulting in the oxidation of NAD from its reduced
form NADH to the oxidized form NAD$^+$. Alterations in redox balance indicating changes in cellular metabolism can be
characterized by the ratio of fractional concentrations of the  oxidized and reduced forms of NAD~\cite{Moreira2016}. In
their pioneering works Chance {\latin et al.} \cite{Chance1962, Barlow1976} have demonstrated that NADH exhibits
autofluorescence in its reduced form, whereas NAD$^+$ is not fluorescent. The reduced form of NAD has two UV absorption
bands with maxima at about 260~nm and 340~nm associated with adenine (AD) and nicotinamide (NA) chromophore groups, respectively
and a fluorescence band with a hump at 460~nm. In turn, the oxidized form NAD$^+$ has only one absorption band peaked at 260~nm and does
not fluoresce. Chance {\latin et al.} \cite{Chance1962, Barlow1976} suggested that the changes occurring in redox state in
cancer cells can be effectively studied by monitoring NADH fluorescence. This approach was developed and widely used
by many groups (see, e.g. \cite{Kasischke2004,Blinova2005}) however further experiments elucidated a
number of biological and technical problems dealing with light scattering in tissue, absorption of the laser light at
360~nm by other intracellular species, and the influence of a number of factors on fluorescence intensity.

To overcome these problems Lakowicz {\latin et al.}~\cite{Lakowicz1992} employed the measurements of NADH fluorescence lifetimes
that are independent of light intensity and can be used for separation of free and protein-bound NADH quantitatively.
Fluorescence lifetime imaging microscopy (FLIM) is now widely used for monitoring of metabolic pathways activated
in cancer cells\cite{Yaseen2017,Evers2018,Schaefer2019}.

As known from the time-resolved studies of fluorescence in cells and living tissues, NADH shows a multicomponent
decay dynamics. In the nanosecond time domain it is most common to fit NADH fluorescence decay after excitation of the
NA chromophore group by two exponentials with lifetimes of about 2.0~ns and 0.4~ns that are usually attributed to the
enzyme-bound and free forms, respectively \cite{Konig1997,Vishwasrao2005}. In aqueous solution free NADH
is also known to exhibit biexponential fluorescence decay with the lifetimes of about 0.3~ns and
0.7~ns~\cite{Visser1981,Couprie1994a,Hull2001}. In other solvents the lifetimes and corresponding weighting
coefficients depend strongly on temperature and solvent type \cite{Ladokhin1995, Couprie1994a, Visser1981,Blacker2013}.
Despite numerous studies carried out and various models suggested the origin of the two fluorescence decay
times still remains controversial. In particular it is unclear if the two fluorescence lifetimes either correspond to different geometric conformations of the whole molecule, or they are intrinsic features of the NA ring.

Although the results of NMR experiments \cite{Oppenheimer1971,McDonald1972} and theoretical studies from \emph{ab initio} point of view \cite{Babu2016,Dolg2019} unambiguously suggest the existence of numerous possible conformations in NADH, several authors attributed all possible conformations into two main groups: folded, where the NA and AD rings are stacked, and unfolded  (more extended) form where the rings are well separated from each other.~\cite{Visser1981,Couprie1994a,Ladokhin1995,Hull2001,Blacker2013,Peon2020} An important role of the NADH conformations in the excited state dynamics was manifested by investigations of direct intramolecular energy transfer
between the AD and NA moieties under excitation within UV absorption band at 260~nm. \cite{Freed1967,Hull2001,Heiner2017,Peon2020} They showed that the efficiency of the energy transfer is high in aqueous solution but decreases rapidly in methanol-water solutions due to denaturation effect. Heiner et al.\cite{Heiner2017} have recently applied ultrafast transient absorption measurements in NADH and reported a very rapid energy
transfer rate of ~70~fs that they explained by a Forster-type mechanism. The experimental transient signals contained also a 1.7~ps decay associated with vibrational and electronic relaxation and a 650~ps decay that was in a good agreement with the known value of long fluorescence decay time in NADH.~\cite{Heiner2017} The study of picosecond vibrational dynamics in NADH first excited state under excitation at 360 nm in water-ethanol solutions  has recently been reported in our recent publications~\cite{Gorbunova20,Gorbunova20b} using a novel polarization-modulation pump-probe transient spectroscopy. Very recently the analysis of spontaneous emission signals from NADH in the sub-picosecond and picoseconds time-scales obtained by means of up-conversion detection scheme was reported by Cao {\latin et al.} \cite{Knutson2019,Knutson2020} and Cadena-Caicedo et al.\cite{Peon2020}. These studies clarified further details of the energy transfer process.

Explaining the nature of two fluorescence decay times Visser {\latin et al.} \cite{Visser1981} suggested that in the folded
molecular conformation the NA and AD rings form an exciplex with a lifetime of about two times longer
than that of the unfolded conformation. Gafni and Brand \cite{Gafni1976} and Ladokhin et al.\cite{Ladokhin1995}
considered the existence of a reversible chemical reaction in NADH excited state resulting in a biexponential
fluorescence decay kinetics. However, the models suggesting a strong relationship between two decay times
observed and the folded and unfolded molecular conformations failed to explain important experimental observations. These are: relatively stable values of the preexponential factors in NADH in solutions of different polarity and existence of two decay times in mononucleotides
NMNH ($\beta$-NA mononucleotide) and MNH (1-methylNA) that lack the AD moiety.
\cite{Krishnamoorthy1987,Kierdaszuk1996,Blacker2019}

Krishnamoorthy {\latin et al.} \cite{Krishnamoorthy1987} and Kierdaszuk et al.\cite{Kierdaszuk1996} suggested that the
heterogeneity in the measured lifetimes arises from the inherent photoprocess of the dihydronicotinamide chromophore
and is not due to intramolecular interactions between the NA and AD parts of NADH. This approach has recently been specificated by
Blacker \emph{et al.}\cite{Blacker2013,Blacker2019} who suggested that the two fluorescent states correspond to alternate
\emph{cis} and \emph{trans} configurations of the NA ring. Blacker\emph{ et al.}\cite{Blacker2013} argued that the
two fluorescence lifetimes observed in NADH depend mostly on non-radiative decay which can be treated as
conformational relaxation by activated barrier crossing and described by the Kramers' hydrodynamic theory and that the
heterogeneity in the non-radiative decay is due to small scale motions such as ring puckering. \cite{Wu1995,Hurley1997}

This paper aims to address further important questions on the NADH excited state dynamics under excitation of the
dihydronicotinamide moiety within the first absorption band that despite of many studies done still remain
controversial. These are: the role of NADH conformations in excited state dynamics, the role of \emph{cis} and
\emph{trans} configurations of the NA ring in the heterogeneity in the measured lifetimes, the influence of
solution viscosity and polarity on the measured decay times and rotation diffusion time.

\begin{figure}[h]\center
\includegraphics[width=0.8\textwidth]{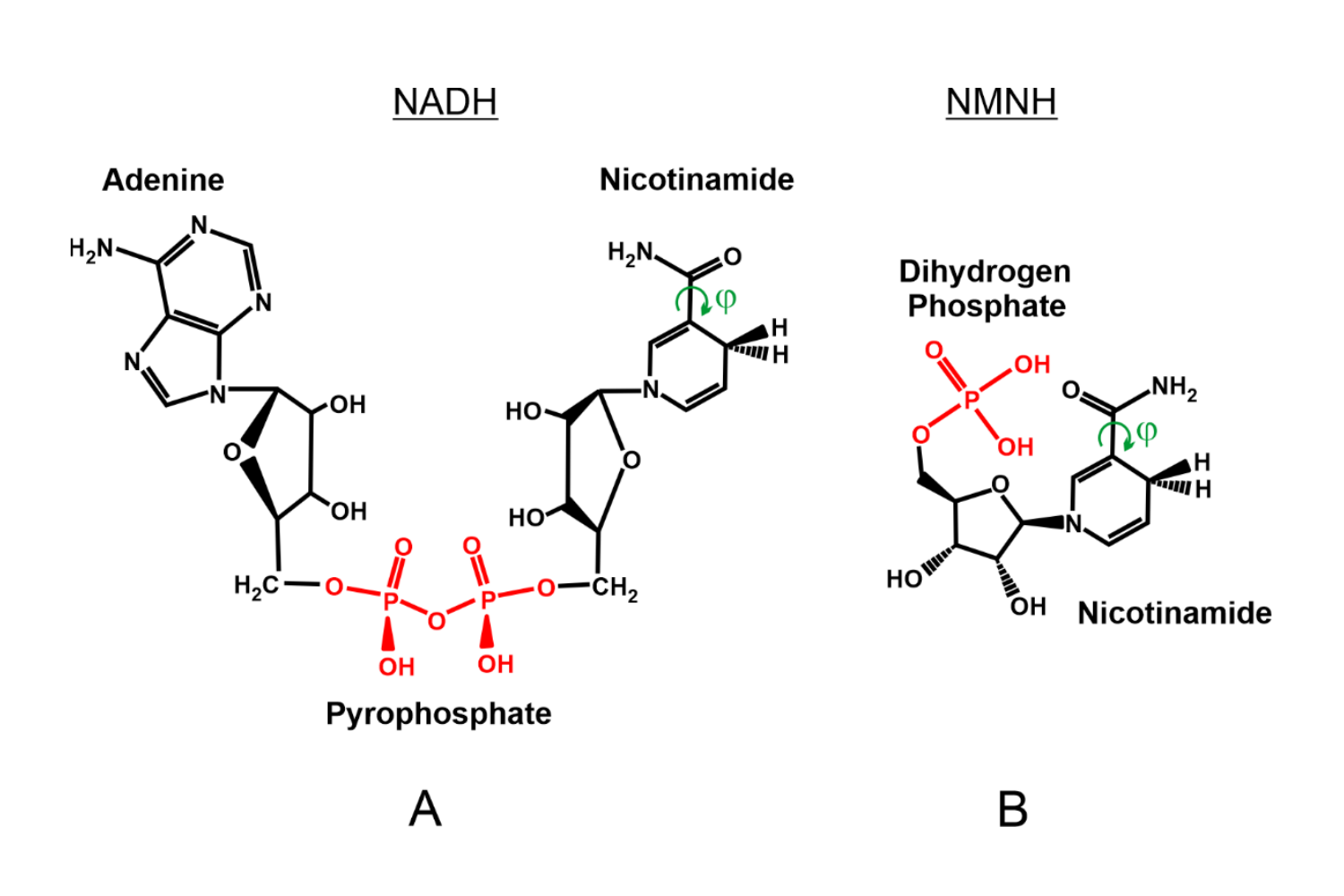}
\caption{Schematic of NADH and NMNH}
\label{fig:NADH}
\end{figure}

We studied polarized fluorescence in NADH at 460~nm under two-photon femtosecond excitation at 720~nm in
water-methanol solutions as a function of methanol concentration and determined a number of fluorescence parameters
using the global fit procedure developed in our previous papers. A comprehensive analysis of the experimental errors
was made. The interpretation of the experimental results obtained was supported by \emph{ab initio} calculations of the
NADH and NMNH structures (see sketches in Fig.\ref{fig:NADH}) in various solutions. The conclusions made were compared with the
results reported by other authors.

In brief, the results obtained and conclusions made are as follows. The consideration of only folded and unfolded conformation groups is insufficient for understanding the excited state dynamics in NADH because interaction between the NA and phosphate moieties are sometimes more important than the interaction between the NA and AD ones (see Fig.1A). The experimental data can be fitted satisfactory by two exponents with the isotropic decay times $\tau_1$ and $\tau_2$ and a single exponents with the rotation diffusion time $\tau_r$: the inclusion of additional exponential decays does not improve the fit quality within given experimental errors. The heterogeneity in the measured decay times in NADH and NMNH can be explained by the existence of different charge distributions of the \emph{cis} and \emph{trans} configurations in the nicotinamide ring that results in different electrostatic field distributions and different non-radiative decay rates. The analysis of the rotational diffusion time $\tau_r$ recorded as a function of methanol concentration on the basis of the Stokes-Einstein-Debye equation allowed for determination of relative concentrations of the folded and unfolded NADH conformations. The recording and analysis of the fluorescence anisotropy parameters and parameter $\Omega$ allowed for determination of the two-photon excitation tensor components and suggested the existence of two excitation channels with comparable intensities: the longitudinal excitation channel dominated by the two-photon diagonal tensor component $S_{zz}$ and the mixed excitation channel dominated by the off-diagonal tensor components $|S^2_{xz}+S^2_{yz}|^{1/2}$.

The organization of the paper is as follows. Section I contains the description of the experimental method used. Section II presents the experimental signals and their analysis. The experimental results obtained and comparison with the results of other authors are given in Section III. The description of the \emph{ab initio} calculations performed are given in Section IV. The discussion of the obtained results and the models developed are given in Section V. The Conclusion summarises the main results obtained. The details of the derivations and the results of \emph{ab initio} calculations are given in \emph{Supporting Information} (\emph{SI}).

\section{I. Experimental method}
\subsection{Detection of polarized fluorescence}
A standard experimental procedure of detection of two-photon excited polarized fluorescence \cite{Lakowicz97a} was used
in experiments. Particular details of the experimental setup used were described in our previous publications.
\cite{Herbrich15,Sasin18,Sasin19} Briefly, two-photon excitation of NADH-containing solutions was performed by a
pulsed laser beam propagated along X axis. The laser beam polarization was either linearly polarized along the vertical,
or horizontal axes, or was left handed circularly polarized. Polarized fluorescence was collected along Z axis at the right angle
to the laser beam propagation. Two orthogonally polarized X and Y fluorescence components were separated by a
Glan prism and then independently detected by two fast photodetectors operated in a photon counting mode. The recorded
signals were analyzed by a time correlated single photon counting (TCSPC) system.

The experimental time-resolved fluorescence polarization components were processed using the expressions \cite{Herbrich15}:
\begin{eqnarray}
\label{eq:y conv}
I_y(t) &=& G\int_{-\infty}^t IRF(t')I_l(t-t')\left[1 + 2r_l(t-t')\right]dt',\\
\label{eq:x conv}	
I_x(t) &=& \int_{-\infty}^t IRF(t')I_l(t-t') \left[1 - r_l(t-t')\right]dt',
\end{eqnarray}
where $I_l(t-t')$ and $r_l(t-t')$ are isotropic intensity and anisotropy, respectively,  $IRF(t')$ is an instrumental response function, and G is a ratio of
sensitivity of X and Y fluorescence detection channels.

The subscript index $l$ in eqs.(\ref{eq:y conv}) and (\ref{eq:x conv}) refers to two possible  linear
polarizations of the laser beam along Y and Z axes. In the case of circular beam polarization the subscript index $l$ should be replaced by index $c$ and subscript indices $x$ and $y$ in
eqs.(\ref{eq:y conv}) and (\ref{eq:x conv}) should be exchanged.

Based on the results of previous experiments (see e.g. \cite{Visser1981,Couprie1994a,Hull2001,Blacker2013,Blacker2019}) a double exponent expression was used in eqs.(\ref{eq:y conv}) and (\ref{eq:x conv}) for the isotropic part of the fluorescence intensity in NADH:
\begin{equation}
\label{eq:iso}
I_l(t-t')=I_l\left(a_1exp\left(-\frac{t-t'}{\tau_1}\right)+a_2exp\left(-\frac{t-t'}{\tau_2}\right)\right)
\end{equation}
where $\tau_1$ and $\tau_2$ are fluorescence decay times, $a_1$ and $a_2$  are corresponding weighting coefficients ($a_1+a_2=1$), and $I_l$ is the time-independent maximum value of the isotropic fluorescence intensity.

The anisotropy $r_l(t-t')$ in eqs.(\ref{eq:y conv}) and (\ref{eq:x conv}) was shousen in a simplest single-exponent form
\begin{equation}
\label{eq:aniso}
    r_l(t-t')=r_l\,exp\left(-\frac{t-t'}{\tau_{r}}\right),
\end{equation}
where  $\tau_{r}$ is the rotation diffusion time.

For circularly polarized excitation light the subscript index $l$ in eqs.(\ref{eq:iso}) and (\ref{eq:aniso}) should be replaced by the index $c$. As shown below, a single-exponential expression in eq.(\ref{eq:aniso}) was sufficient for obtaining the fit quality consistent with the experimental errors available. The fluorescence parameters $I_l$, $I_c$, $r_l$, $r_c$, $\tau_1$, $\tau_2$, $a_1$, $a_2$, and $\tau_r$ were fitted from experimental data using eqs.(\ref{eq:y conv}) and (\ref{eq:x conv}).

\subsection{Materials}
Reduced $\beta$-NADH (Sigma–Aldrich) was used in experiments. Distilled water and methanol of 96\% purity were used for
preparing solutions at volume proportions ranging from pure water to 100\% methanol. The NADH concentration in all
cases was the same and equal to 1~mM. All solutions were prepared fresh daily and used at the temperature of 20$^{\circ}$C.

\subsection{Experimental procedure}
A femtosecond Ti:Sa oscillator (Mai Tai HP, Spectra Physics) tunable in the spectral range of 690–1040~nm with a pulse
duration of 100~fs and a repetition rate of 80.4~MHz was used as an excitation source. About 2\% of the laser output
at 720~nm was split out and sent to the synchronization channel consisted of an attenuator, a lens, and a fast silicon
photodiode for synchronization of the fluorescence signals in the TCSPC module. The output laser beam was attenuated
and then expanded by a telescope to the diameter of 4~mm. The laser beam polarization was controlled by a half-, or
quarterwave plates. The  linear polarization degree of the laser beam was better than 0.995, and the circular polarization
degree was better than 0.95. The polarized laser beam was focused onto the center of a quartz cuvette containing NADH
in water-methanol solution. Average power of the laser beam on the cuvette was kept at the level of about 100~mW.

The fluorescence was collected in the direction perpendicular to the laser beam propagation and formed by a lens into a
quasi-parallel beam that was split by a Glan prism into two beams with orthogonal polarizations parallel to X and Y
axes. Two orthogonal polarization components were detected by ultrafast avalanche photodiodes (ADP-050-CTC, MPD) with a
HWHM of instrumental response function of about 56~ps. The IRF contour was determined experimentally by recording laser pulses scattered from a fiber sample. As found the IRF shape in the conditions of our experiments had a strong spectral dependence, therefore the spectral range of the detected NADH fluorescence  was restricted by a narrow-band interference filter with the bandwidth of 460–-490~nm installed after the cuvette in front of the Glan prizm. The fluorescence signals were analyzed by a TCSPS module (Picoharp 300, Picoquant). The fluorescence photons were counted for 350~s for each laser beam polarization with a time bin of 4~ps.

\section*{II. Experimental signals and data analysis}
\subsection*{Experimental signals}
\begin{figure}[h]
\centering
\includegraphics[scale=0.6]{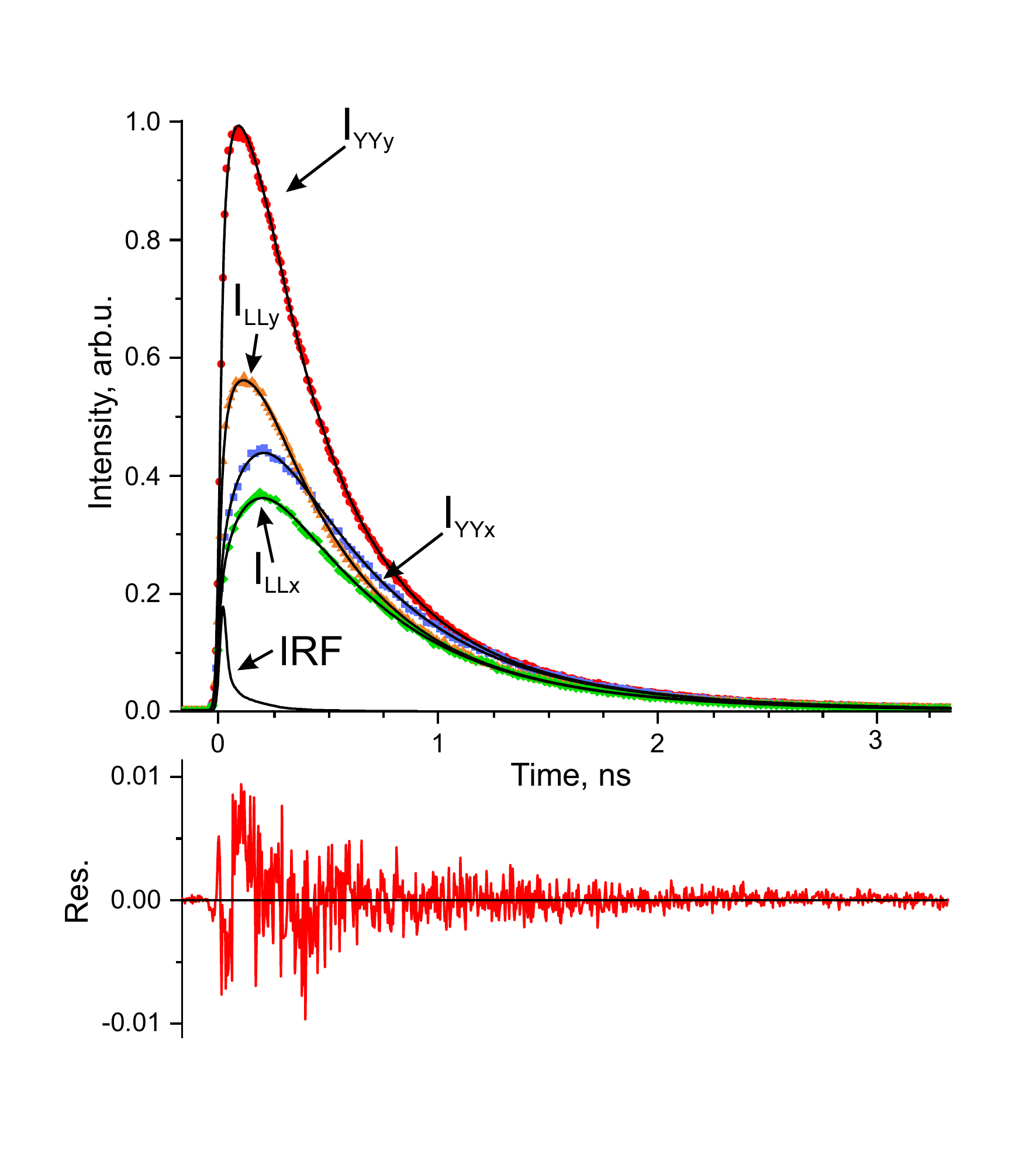}
\caption{Fluorescence decay signals in NADH in aqueous solution under two-photon excitation at 720~nm. \\
Four different data sets relate to different combinations of the laser and fluorescence polarizations, see text for details. Symbols represent experimental data and solid curves are fits. }
\label{fig2}
\end{figure}

According to the results of our recent research~\cite{Sasin19} the fluorescence parameter values  were independent of the excitation wavelength in the range 720 - 780~nm within the experimental error bars. Therefore, in this study the excitation wavelength was fixed at 720~nm.  Typical fluorescence experimental signals in NADH under two-photon excitation  in aqueous solutions for several combinations of polarization of the pump and fluorescence beams are shown in Fig.\ref{fig2}.

The abbreviation $\mathbf{I}$ in Fig.\ref{fig2} with subscripts is the fluorescence intensity, where capital case subscript indices $YY$ and $LL$ label the polarization of pumping photons: $Y$ and $L$ denote linear polarization along axis Y and left handed circular polarization, respectively. The lower case subscripts $x$ and $y$  label the fluorescence beam polarizations. The initial laser beam intensity was the same for all four data sets in Fig.\ref{fig2}, therefore different amplitudes and shapes of the four decay curves indicate the dependence of the two-photon excitation probability and fluorescence decay dynamics on the pump and fluorescence polarization indices.

\subsection*{Experimental data fitting and estimation of experimental errors}
Processing of experimental signals and determination of the fluorescence parameters based on eqs.(\ref{eq:y conv}) and (\ref{eq:x conv}) was carried out by a global fit procedure using a least square differential evolution fitting algorithm with convolution. Experimental data related to $x$ and $y$ fluorescence decay polarization components similar to that shown in Fig.\ref{fig2} were fitted using eqs.(\ref{eq:y conv}) and (\ref{eq:x conv}) separately for linearly and circularly polarized excitation light aimed to determine the global fluorescence parameters $I_l$, $r_l$, $\tau_1$, $\tau_2$, $a_1$, $a_2$, $\tau_r$ and $I_c$, $r_c$, $\tau_1$, $\tau_2$, $a_1$, $a_2$,  $\tau_r$, respectively. As a result of these fits the parameters in each set responsible for fluorescence dynamics at the time $t>0$: $\tau_1$, $\tau_2$, $a_1$, $a_2$, and  $\tau_r$  always had close values within experimental error bars. These parameters were assumed to be the same in both sets and their values were averaged to obtain final values.

Different sources of experimental data errors were carefully analyzed. According to the analysis made, a statistical method of error estimation dealing with the standard deviation by optimization of the cost function near the global minimum in the case of fitting with exponential functions usually resulted in significant underestimation of actual errors because the statistical errors obtained were typically much smaller than systematic experimental errors.

Although the fit quality obtained when processing the experimental data often looked perfect with the naked eye, the values of the fluorescence parameters fluctuated from one experiment to another much more than was given by the statistical analysis of each data set. To take proper account for systematic experimental errors the experiments were repeated several times at the same conditions to obtain several experimental data sets (usually six $x$ and $y$ pairs of fluorescence decay data), and  the fluorescence parameters were calculated by fit for each set separately.  Then the parameter mean values and standard deviations were calculated by averaging of all sets values with 95 \%  confidence  using  Student's $t$-value. The error bars determined in this way are used throughout this paper.

The role of the IRF($t$) shape shown in Fig.\ref{fig2} on the determined fluorescence parameter values  was carefully analyzed and found to be very important in the conditions of our experiment especially for determination of the rotational diffusion time $\tau_r$ and weighting coefficients $a_1$ and $a_2$.  Even small perturbations at the rising edge and the tail of IRF led to noticeable changes in the calculated values of the fluorescence parameters. Several IRF functions were used for experimental data analysis: experimentally determined IRF in a numerical form, smoothed experimental IRF, and analytical IRF form obtained by approximation of the experimental IRF with two Gaussian functions. The best results were obtained by using the experimental IRF smoothed without deformations of the rising edge and the tail. This IRF was used for analysis of the experimental data.
The details of the fitting protocol and analysis of statistical and experimental errors are given in \emph{SI}.

\subsection*{Experimental data analysis}
The analysis of the obtained fluorescence parameters was based on the general expression for fluorescence intensity $I(t)$ after two-photon excitation derived in our previous publications \cite{Shternin10,Denicke10} and given in eq.(\ref{eq:intens_final}) in \emph{SI}.

The expression for the fluorescence intensity in eq.(\ref{eq:intens_final}) was similar to that given earlier by McClain \cite{McClain73} and Wan and Johnson \cite{Wan94} however it was presented in a somehow more compact
spherical tensor form allowing for complete separation of the light polarization part controlled by experimentalist from the molecular dynamics part described by the molecular parameters $M_{K_e}(R,R',t)$ ($M$-parameters).

The $M$-parameters  are real values that can be extracted from experiment and contain all information on the symmetry and structure of the molecular electronic transitions. The set of $M$-parameters at the time of excitation $t=0$ is equivalent to the set of McClain's Cartesian
parameters $\hat Q_i$. The relationship between the two sets of parameters is tabulated in ref. \cite{Shternin10}. In the conditions of the collision-induced rotational diffusion the $M$-parameters can be expressed
as~\cite{Shternin10}:
\begin{equation}
\label{eq:MK}
M_{K_e}(R,R',t)=
-\sqrt{3}\sum_{q_e,q'_e}\left(\left[{\sf
\textbf{F}_1^*}\otimes {\sf \textbf{F}'_1}\right]_{K_eq_e} {\cal
D}^{K_e}_{q_e,q'_e}(t)\left[{\sf \textbf{S}^*}_{R'}\otimes {\sf
\textbf{S}}_R \right]^*_{K_eq'_e}\right),
\end{equation}
where the symbol $\otimes$ denotes tensor product\cite{Zare88b} and ${\cal D}^{K_e}_{q_e,q'_e}(t)$, where $K_e=0,2$, is a rotational diffusion tensor matrix element given in eq.(\ref{eq:D}) in \emph{SI}. If $K_e=0$ this matrix element is equal to unity and if $K_e=2$ in the condition of our experiment it is proportional to:
\begin{equation}
\label{eq:D00}
{\cal D}^{2}_{q_e,q'_e}(t) \sim \delta_{q_e,0}\delta_{q_e',0}e^{-t/\tau_r},
\end{equation}
where $\tau_r$ is the rotational diffusion time.

The  spherical components of the two-photon excitation tensor $\textbf{S}_{R\gamma}$ in eq.~(\ref{eq:MK}) with a rank $R$ and its component $\gamma$ are given by~\cite{Shternin10}:
\begin{eqnarray}
\label{eq:SR}
 S_{R\gamma}&=&2\sum_{q_1, q_2, n_i} C^{R\gamma}_{1q_1\;1q_2}\frac{
 \langle n_e | \hat{d}_{q_2} | n_i \rangle\langle n_i| \hat{d}_{q_1}|
 n_g\rangle}{E_i-E_g-h\nu},
\end{eqnarray}
where $R=0,2$, $\gamma=-R \cdots R$, $\nu$ is a laser photon frequency, the terms in the angular brackets are dipole transition matrix elements, and summation is performed over the molecular frame spherical projections $q_1,q_2=-1,0,1$ and over the quantum numbers of all virtual intermediate states $n_i$.

The terms $E_i$ and $E_g$ in eq.(\ref{eq:SR}) are the electronic energies of the intermediate and ground states, respectively. The body-frame spherical components of the one-photon  emission dipole moment
$\mathbf{F}_1$ in Eq.~(\ref{eq:MK}) are given by:
\begin{eqnarray}
\label{eq:Fq}
{F}_{1 q_{fl}}=\langle n_{f}|{\hat{d}_{q_{fl}}}|\widetilde{n}_e\rangle,
\end{eqnarray}
where $q_{fl}=0,\pm1$ are spherical vector projections, $|\widetilde{n}_e\rangle$ is the relaxed excited electronic state, and $|n_{f}\rangle$ is the final (ground) electronic state.

In the case of a one-color two-photon population of a nondegenerate excited electron state that is relevant to our experimental conditions, the rank $K_e$ in Eq.~(\ref{eq:MK}) is limited to $K_e=0,2$ and the total number of the $M$-parameters is four \cite{Callis.1993}. There are two zeroth rank ($K_e=0$) $M$-parameters: $M_0(0,0)$ and
$M_0(2,2)$ that can have only non-negative values and contribute to the isotropic part of the fluorescence intensity
and two second rank M-parameters ($K_e=2$): $M_2(0,2,t)$, and $M_2(2,2,t)$ that govern the anisotropic,
polarization-dependent part of the fluorescence intensity. The evolution of the $M$-parameters with $K_e=2$ in time
due to the rotational diffusion is expressed in Eq.~(\ref{eq:D00}) while the zeroth-rank parameters do not
depend on time.

All four parameters can be directly determined from experiment (see {\latin e.g.} \cite{Denicke10}) by combining different laser and fluorescence polarizations. The $M$-parameters can be readily expressed in terms of the anisotropy in eq.(\ref{eq:aniso}) and the parameter $\Omega$\cite{Lakowicz97a} as follows:
\begin{eqnarray}
\label{eq:rl}
r_l(t)=\frac{2}{\sqrt{35}}\frac{\sqrt7 M_2(0,2,t)+M_2(2,2,t)}{\sqrt5 M_0(0,0)+2M_0(2,2)}\,,
\end{eqnarray}
\begin{eqnarray}
\label{eq:rc}
r_c(t)=-\frac{1}{\sqrt{35}}\frac{M_2(2,2,t)}{M_0(2,2)}\,,
\end{eqnarray}
\begin{eqnarray}
\label{eq:Omega}
\Omega=\frac{I_{LL}}{I_{YY}}=\frac{3M_0(2,2)}{\sqrt5 M_0(0,0)+2M_0(2,2)}.
\end{eqnarray}

The M-parameters are very important because according to eqs.~(\ref{eq:MK}), (\ref{eq:SR}), and (\ref{eq:Fq}) they contain all information on the dynamics of the two-photon excitation in terms of the quantum mechanical transition amplitudes and phases. The M-parameters are more general than the components of the two-photon absorption tensor $\mathbf{S}$ in eq.~(\ref{eq:SR}) because they are scalar quantities that do not depend on the
particular excitation model and on the choice of the coordinate frame. The $S_{R\gamma}$ values in eq.~(\ref{eq:SR}) can be determined from experiment under certain assumptions, in particular one should know the direction of the fluorescence transition dipole moment $\mathbf{F}_1$ in the relaxed exited state.

\section{III. Experimental Results}
\begin{figure}[h]
\centering
\includegraphics[scale=0.5]{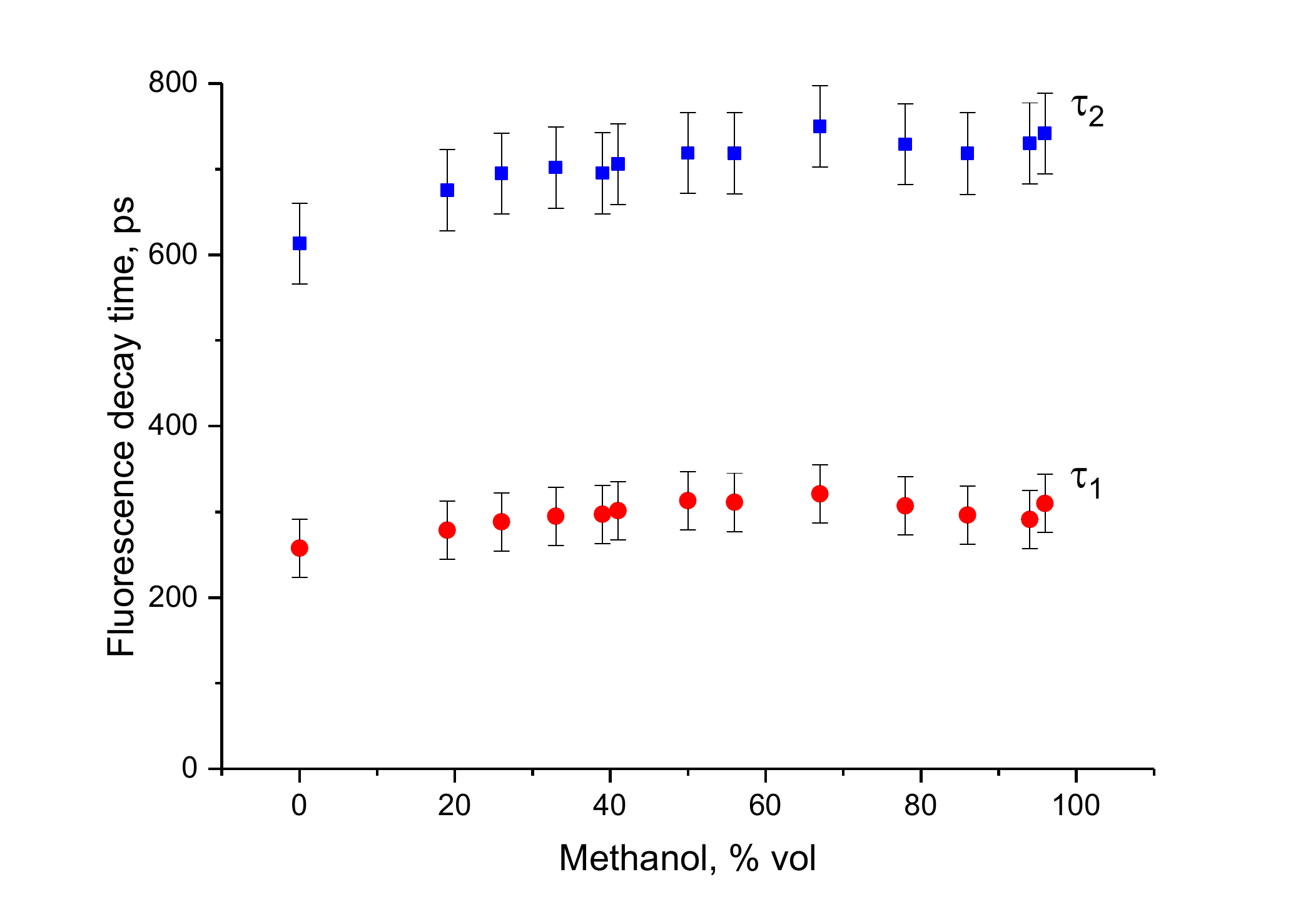}
\caption{Decay times $\tau_1$ and $\tau_2$ in NADH as function of methanol concentration.}
\label{fig:times}
\end{figure}

\begin{figure}[h]
\centering
\includegraphics[scale=0.5]{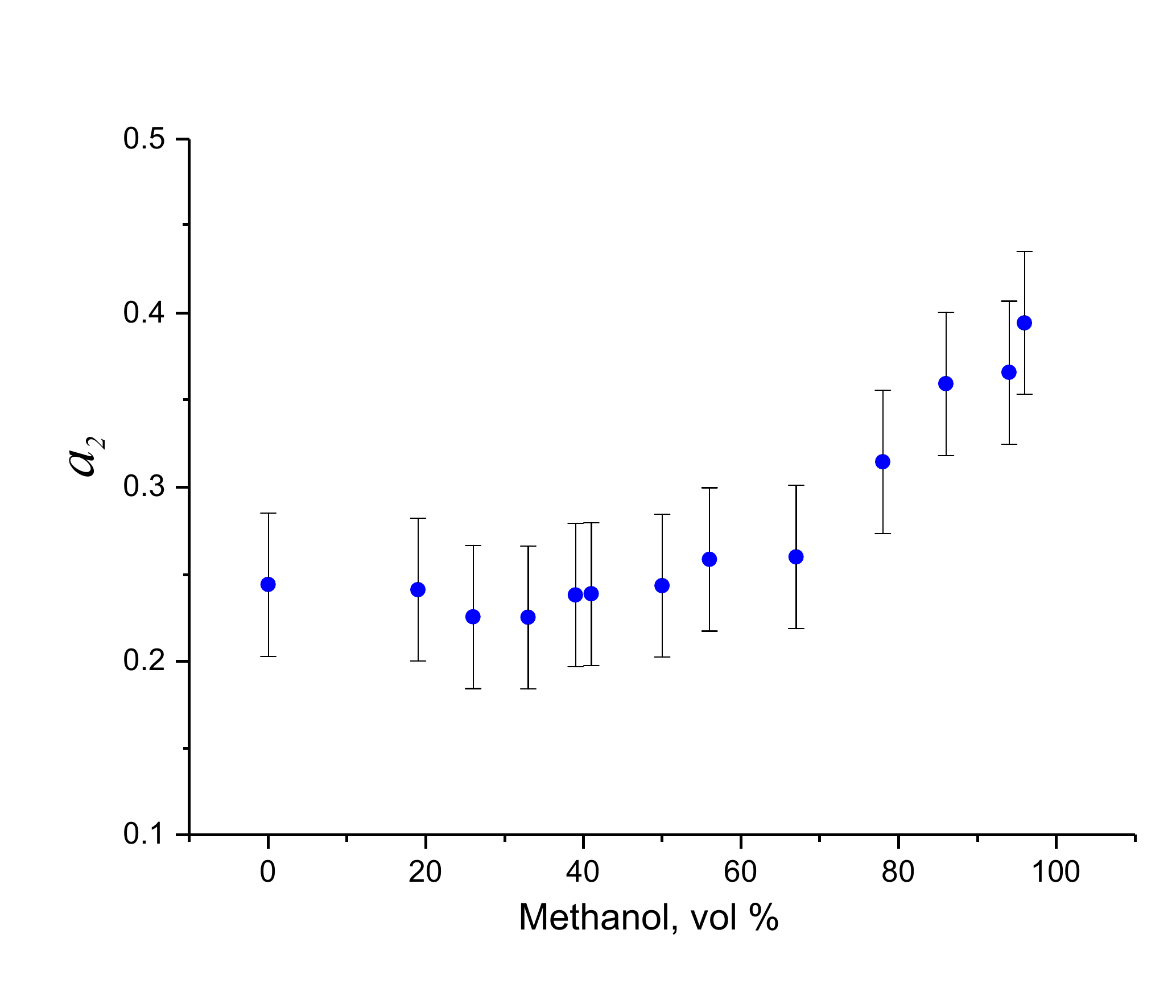}
\caption{Preexponential coefficient $a_2 $ in NADH (see eq.(\ref{eq:iso})) as a function of methanol concentration.}
\label{fig:ratio}
\end{figure}

\begin{figure}[h]
\centering
\includegraphics[scale=0.5]{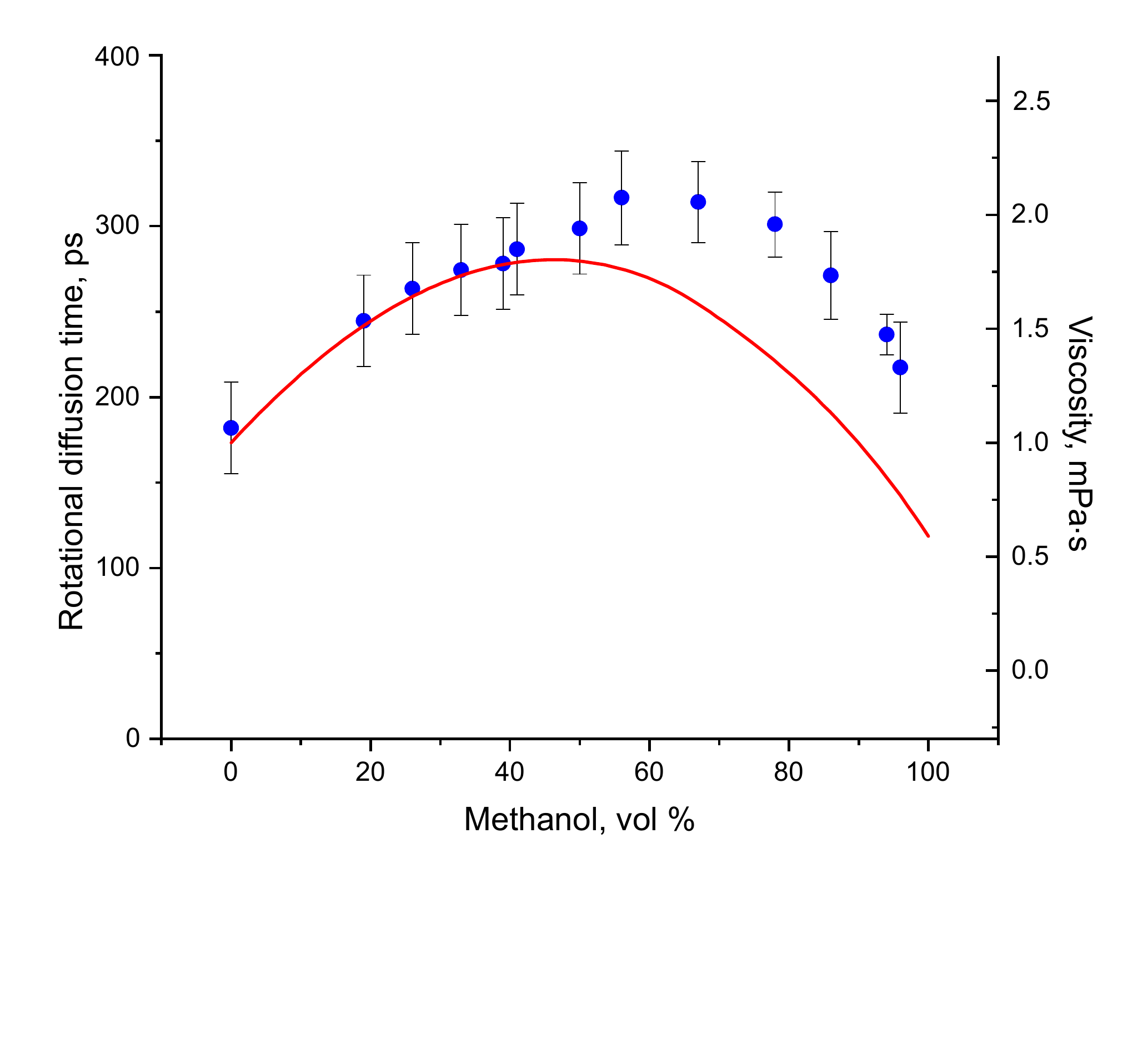}
\caption{Rotational diffusion time $\tau_r$ in NADH as function of methanol concentration. \\
Red solid curve represents solution viscosity.~\cite{Thompson2006}}
\label{fig:rot}
\end{figure}

\begin{figure}[h]
\centering
\includegraphics[scale=0.5]{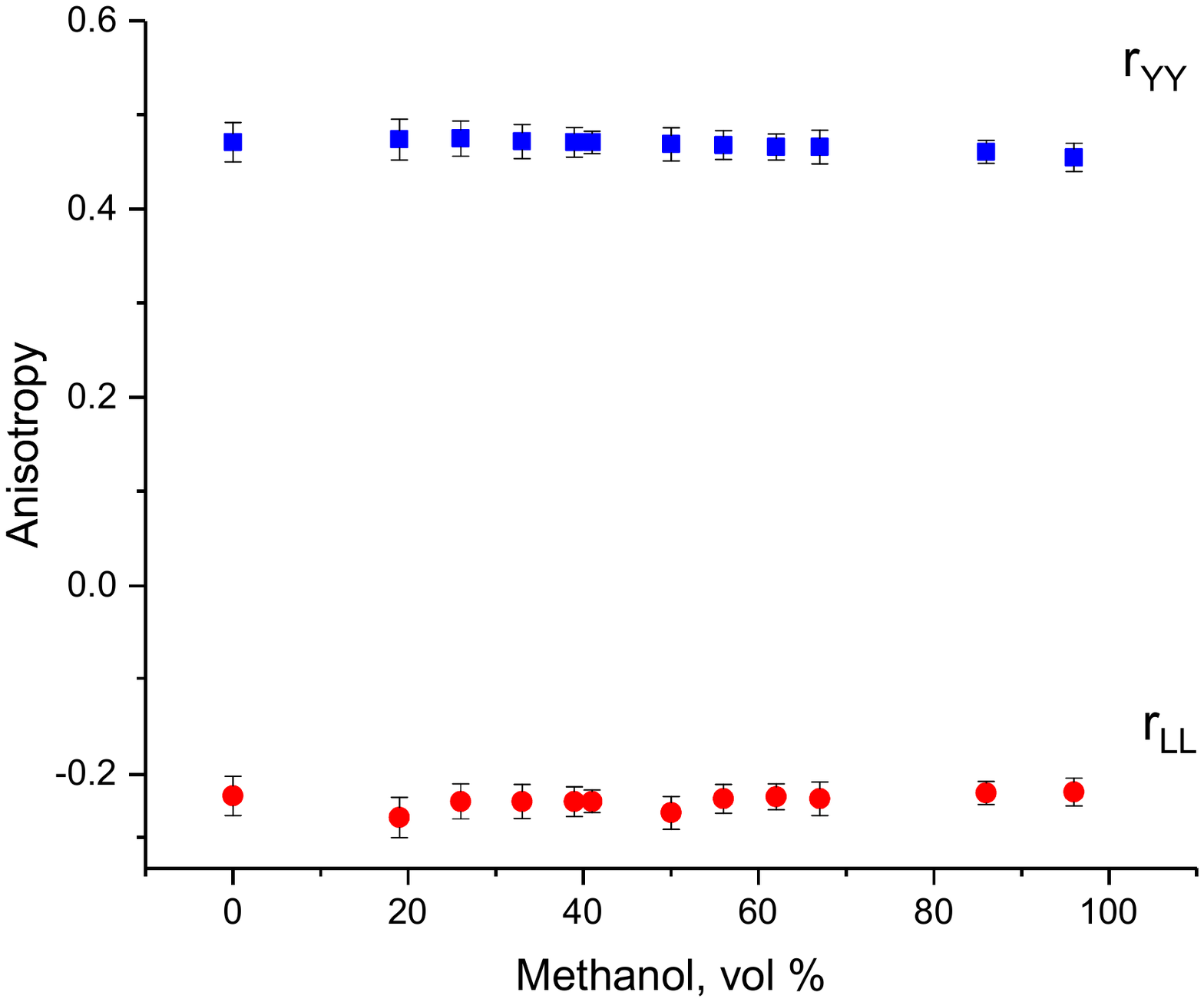}
\caption{Anisotropies $r_l=r_{YY}$ and $r_c=r_{LL}$ under two-photon excitation in NADH as function of methanol concentration. }
\label{fig:anisotropy}
\end{figure}

\begin{figure}[h!]
\centering
\includegraphics[scale=0.5]{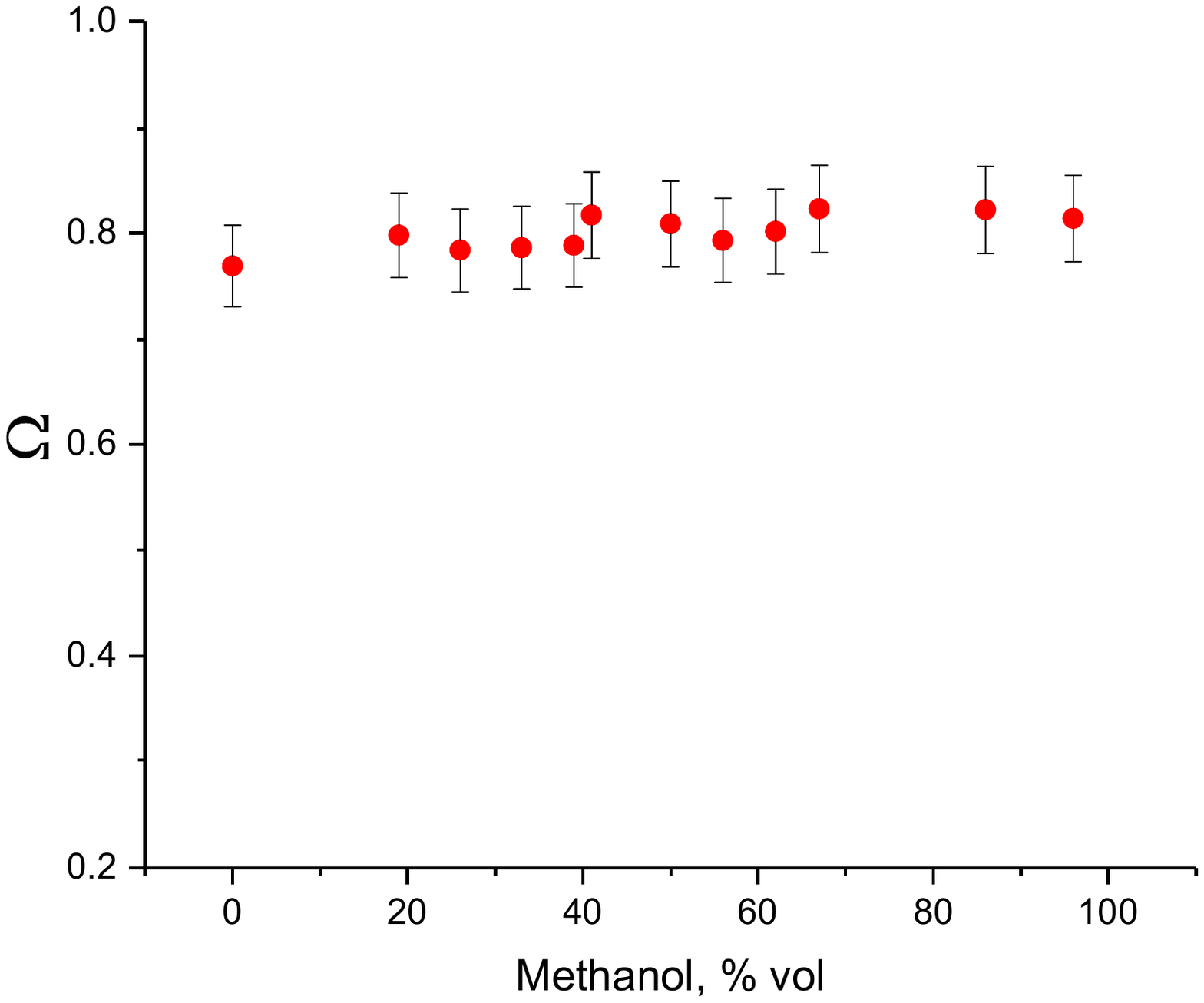}
\caption{Parameter $\Omega$ in NADH as function of methanol concentration. }
\label{fig:Omega}
\end{figure}

The fluorescence decay parameters determined from experiment as a function of methanol concentration are presented in Figs. \ref{fig:times} - \ref{fig:Omega}.

The fluorescence decay times $\tau_1$ and $\tau_2$ are shown in Fig.\ref{fig:times}. As can be seen in Fig.\ref{fig:times}  both fluorescence decay times increase slightly with methanol concentration at the concentrations below 40\% and remain practically the same within experimental error bars at higher concentrations. In pure aqueous solution the fluorescence decay time $\tau_1$ and $\tau_2$ values in Fig.\ref{fig:times} are in agreement with those reported earlier in NADH  under one- \cite{Couprie1994a,Ladokhin1995} and two-photon \cite{Vasyutinskii2017,Blacker2019} excitation within experimental error bars.  Also, both decay times in Fig.\ref{fig:times} agree  roughly with those reported for NADH in pure methanol~\cite{Krishnamoorthy1987,Ladokhin1995}, where however a three-exponential model was used.

The preexponential coefficient  $a_2 = 1-a_1$ (see eq.(\ref{eq:iso})) is presented in Fig.\ref{fig:ratio}. As can be seen in this figure at methanol concentrations below 70\% the coefficient is almost constant and equal to $a_2 = 0.24$ within the experimental error bars, however at higher concentrations it rises up to 0.4. Note that the coefficient $a_2$ refers to the longer decay time in Fig.~\ref{fig:times}.

At zero  and 50\% methanol concentrations the coefficient $a_2 $ values in Fig. \ref{fig:ratio} are in agreement  with those reported by Blacker \latin{et al}.\cite{Blacker2019}, with our recent results \cite{Sasin19,Vasyutinskii2017}, and with the earlier results of  Krishnamoorthy \latin{et al.} \cite{Krishnamoorthy1987} within experimental error bars. In NADH in pure methanol the $a_2$ value in Fig. \ref{fig:ratio} is in general agreement with the result reported by Krishnamoorthy \latin{et al.} \cite{Krishnamoorthy1987} who however used a three-exponential decay fit with a minor ($a_3=0.15$) contribution from the third exponent with $\tau_3=2.0$~ns. Ladokhin \latin{et al.}  \cite{Ladokhin1995} used two, three, and four-exponential decay fits and also reported for NADH diluted in pure methanol the increase of a relative contribution from the longer $\tau_2$ time. However, {quantitatively} the results of Ladokhin \latin{et al.} \cite{Ladokhin1995} differ both from the results shown in  Fig. \ref{fig:ratio} and from those reported by Krishnamoorthy \latin{et al.} \cite{Krishnamoorthy1987} We do not know the exact reason for this discrepancy, however according to our analysis the coefficients $a_1$ and $a_2$ are the most variable among all fluorescence parameters as they can noticeably change their values even at small variations of the experimental conditions and fitting procedure details.

The rotational diffusion time $\tau_r$ is presented in Fig.~\ref{fig:rot}. The water-methanol solution viscosity~\cite{Thompson2006} is also shown in this figure by a red solid curve. One can see that the viscosity in Fig.~\ref{fig:rot} is represented with a non-linear curve with a maximum at about 40\% of methanol. As can be seen in Fig. \ref{fig:rot}, at methanol concentrations from zero to 40\% the rotational diffusion time $\tau_r$ raises proportionally to the solution viscosity, has a maximum at about 60\% of methanol and then decreases.  However at methanol concentrations over 40\%  the direct proportionality between the  $\tau_r$ and the viscosity is violated.

As shown in Fig. \ref{fig:rot} in aqueous solution $\tau_r = 180\pm 30$~ps that agrees well with our recent result \cite{Vasyutinskii2017} and with the result reported earlier by Couprie \latin{et al.} \cite{Couprie1994a}  However this value about 1.6 times smaller than the values reported recently by Blacker \latin{et al.}\cite{Blacker2019} at several excitation wavelengths. We believe that this discrepancy could arise because Blacker \latin{et al.}\cite{Blacker2019} probably did not take into account IRF in their fitting procedure.

The initial anisotropies $r_l$ and $r_c$ are presented in Fig. \ref{fig:anisotropy} as function of methanol concentration.  As can be seen in Fig. \ref{fig:anisotropy} both anisotropies practically did not depend on methanol concentration. The anisotropy values are in perfect agreement with  previously reported results \cite{Vasyutinskii2017,Blacker2019}.

The fluorescence parameters determined:  $a_2$,  $\tau_1$,  $\tau_2$,  $r_l$,  $r_c$, $\tau_{r}$, and $\Omega$ are collected in Table \ref{tbl:1}. The numbers in parentheses in Table \ref{tbl:1} are experimental errors.
\begin{table}
  \caption{Experimentally determined fluorescence parameters: NADH in water-methanol solutions.}
  \label{tbl:1}
\begin{tabular}{cccccccc}
\hline
MeOH & $a_2=(1-a_1)$ & $\tau_1$, ps & $\tau_2$, ps & $r_l$ & $r_c$ & $\tau_{r}$, ps & $\Omega$  \\
\hline
 0 \% & 0.24 (0.05) & 260 (30) & 610 (30) & 0.47 (0.02) & -0.22 (0.02) & 180 (30) & 0.77 (0.04)  \\
 19 \% & 0.24 (0.05) & 280 (30) & 680 (30) & 0.47 (0.02) & -0.25 (0.02) & 240 (30) & 0.80 (0.04) \\
 26 \% & 0.22 (0.05) & 290 (30) & 690 (30) & 0.46 (0.02) & -0.23 (0.02) & 260 (30) & 0.78 (0.04)\\
 33 \% & 0.22 (0.05) & 280 (30) & 700 (30) & 0.47 (0.02) & -0.23 (0.02) & 270 (30) & 0.79 (0.04)\\
 39 \% & 0.24 (0.05) & 300 (30) & 700 (30) & 0.47 (0.02) & -0.23 (0.02) & 280 (30) & 0.79 (0.04) \\
 41 \% & 0.24 (0.05) & 300 (30) & 710 (30) & 0.47 (0.01) & -0.23 (0.01) & 290 (30) & 0.82 (0.04)\\
 50 \% & 0.25 (0.05) & 310 (30) & 720 (30) & 0.47 (0.02) & -0.24 (0.02) & 300 (30) & 0.81 (0.04)\\
 56 \% & 0.26 (0.05) & 310 (30) & 720 (30) & 0.47 (0.02) & -0.23 (0.02) & 320 (30) & 0.79 (0.04)\\
 67 \% & 0.26 (0.05) & 320 (30) & 750 (30) & 0.47 (0.01) & -0.22 (0.01) & 310 (30) & 0.80 (0.04)\\
 78 \% & 0.32 (0.05) & 310 (30) & 730 (30) & 0.47 (0.02) & -0.23 (0.02) & 300 (20) & 0.82 (0.04)\\
 86 \% & 0.36 (0.05) & 300 (30) & 720 (30) & 0.46 (0.01) & -0.22 (0.01) & 270 (30) & 0.82 (0.04)\\
 94 \% & 0.37 (0.05) & 290 (30) & 730 (30) & 0.46 (0.02) & -0.22 (0.02) & 240 (20) & 0.81 (0.04)\\
 96 \% & 0.39 (0.05) & 310 (30) & 740 (30) & 0.45 (0.02) & -0.22 (0.02) & 217 (30) & 0.82 (0.04)\\
\hline
\end{tabular}
\end{table}

\newpage

The two-photon excitation parameter $\Omega=I_{LL}/I_{YY}$\cite{Lakowicz97a} in NADH  is presented in Fig. \ref{fig:Omega}. As can be seen the parameter practically does not depend on methanol concentration.

\newpage

\section{IV. Theoretical studies}
\label{S:ab_initio}
\emph{Ab initio} calculations of several  conformers of NADH and of NMNH in vacuum, and in water and methanol solutions in the ground and excited states have been carried out to extract information on molecular equilibrium geometries, electronic energy levels, vertical excitation energies, oscillator strengths, and dispersion energies.

Earlier, \emph{ab initio} calculations of NADH conformers have been reported by Wu \emph{et al.}\cite{Wu1993} at the MP2/6-31G*//6-31G* level and
by Kumar \emph{et al.}\cite{Kumar2010} at the B3LYP/6-311++G** level. To the best of our knowledge, no calculations of NADH dissolved in water and methanol have been carried out till now.

In this paper electronic structure computations of twenty four NADH conformers, including $\textit{cis}$ and $\textit{trans}$ modifications dissolved in water and methanol have been performed \emph{ab initio} by means of the polarizable continuum model (PCM) at the B3LYP-D3BJ/6-31G* level with the GAUSSIAN package~\cite{GAUSSIAN09}. The functionals were extended with the D3 version of Grimme's dispersion correction.~\cite{Kovacs2017}  One of the conformers (N8, see below) has been calculated at the B3LYP-D3BJ/6-31+G* level + (PCM) for comparing parameters.

\begin{figure}[h]
\centering
\includegraphics[scale=0.8]{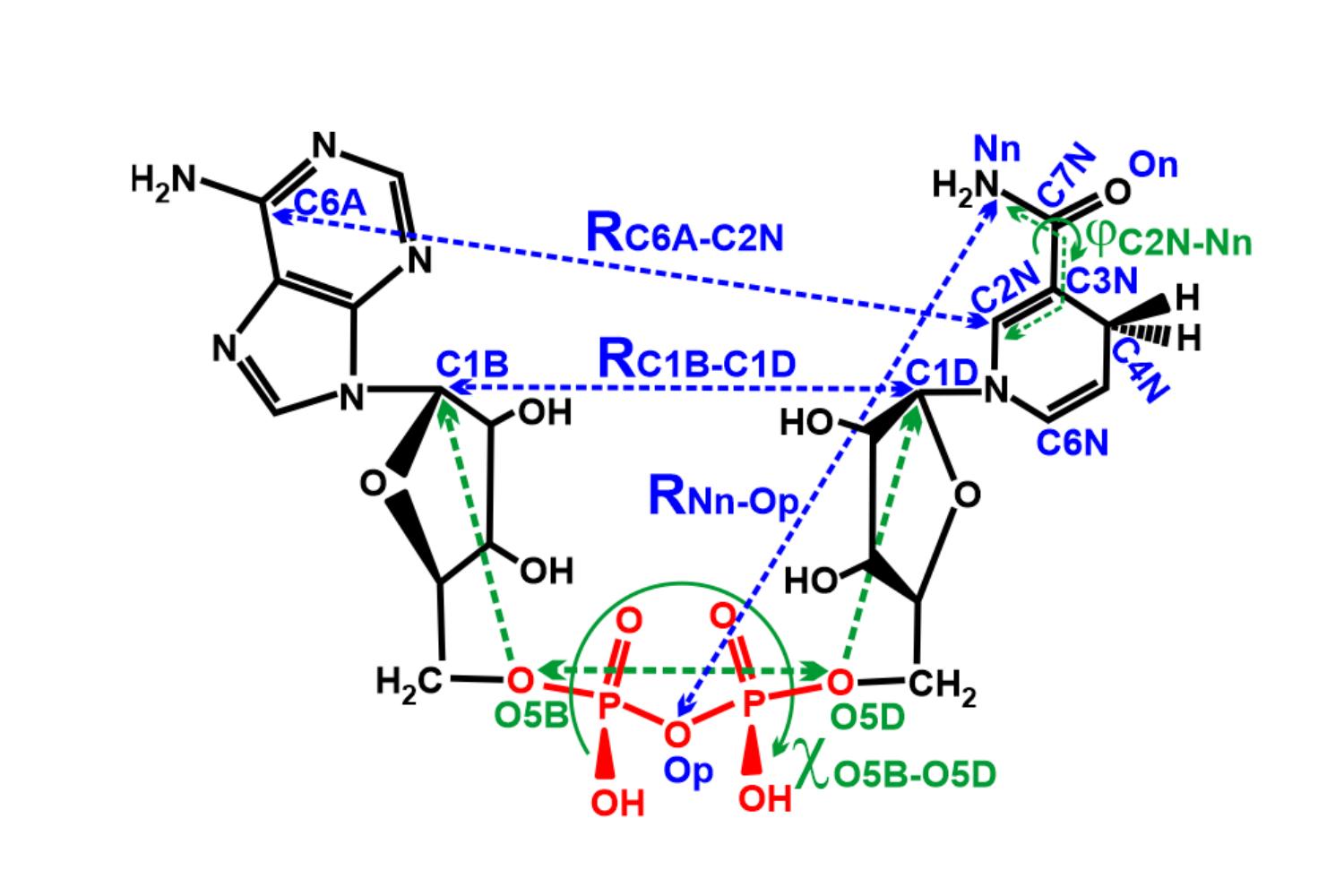}
\caption{Schematic of NADH with important interatomic distances and torsion angles indicated. }
\label{fig:NADHdist}
\end{figure}

A   schematic of NADH is shown in Fig.\ref{fig:NADHdist} where a number of distances and angles between most important atoms and molecular groups are indicated.

Optimized stable ground state geometries in terms of the interatomic distances and torsion angles shown in Fig.\ref{fig:NADHdist} in water and methanol are presented in \emph{SI} in Tables~\ref{tab:nadh_conf_geom_water} and \ref{tab:nadh_conf_geom_methanol}, respectively. The optimized geometries of the first excited state  in aqueous solution are presented in \emph{SI} in Tab.~\ref{tab:nadh_conf_geom_water1}.

Vertical excitation energies of four lowest electron excited states, corresponding one-photon oscillator strengths, and ground state dispersion energies ($E_D$) for $\textit{cis}$ and $\textit{trans}$ forms of twelve NADH conformations in water and methanol solutions are presented in \emph{SI} in Tables~\ref{tab:nadh_ex_energy_water} and \ref{tab:nadh_ex_energy_methanol}, respectively.   According to the vertical excitation energy values in Tables~\ref{tab:nadh_ex_energy_water} and \ref{tab:nadh_ex_energy_methanol} only the first excited state  could be two-photon excited in NADH at 720~nm in the conditions of our experiments. Data on the higher excited states are also presented in both tables and can be used for analysis of interactions between excited molecular moieties.

\begin{figure}[h]
\centering
\includegraphics[scale=0.8]{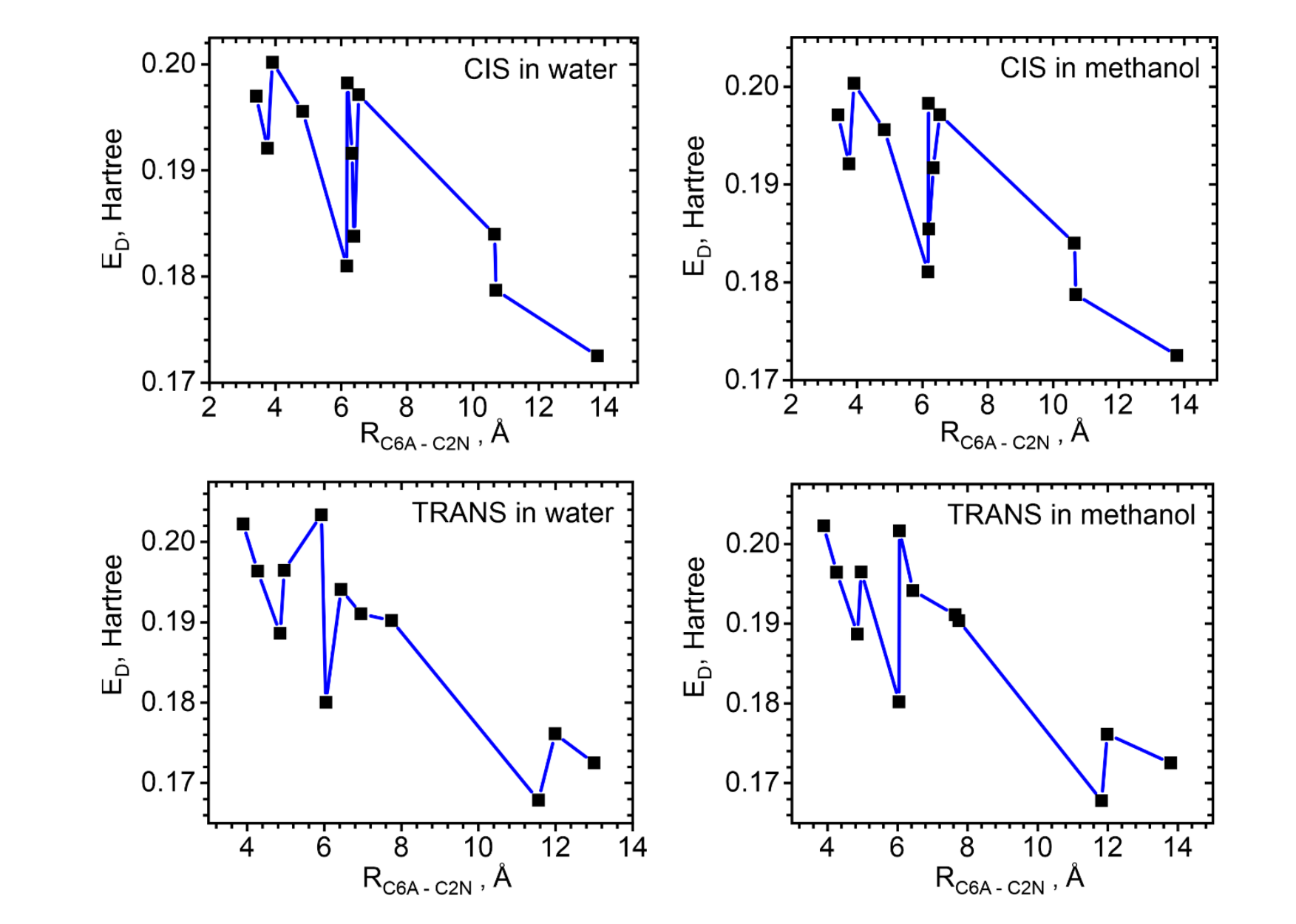}
\caption{Dispersion energy in NADH \emph{cis} and \emph{trans} conformations dissolved in water and methanol as function of the distance
$R_{C6A-C2N}$ in Fig.\ref{fig:NADHdist} }
\label{fig:dispersion}
\end{figure}

As can be seen in Tables~\ref{tab:nadh_ex_energy_water} and \ref{tab:nadh_ex_energy_methanol} in \emph{SI} vertical excitation energies and oscillator strengths  to the first excited state shown in the first columns in the tables vary somewhat from one NADH conformation to another. However they do not show a  distinct correlation with the values of geometrical molecular characteristics in Tables~\ref{tab:nadh_conf_geom_water} and \ref{tab:nadh_conf_geom_methanol}. On the other hand the dispersion energy in the last columns in Tables~\ref{tab:nadh_ex_energy_water} and \ref{tab:nadh_ex_energy_methanol} changes dramatically with respect to the interatomic distances $R_{C6A-C2N}$, $R_{Nn-Op}$ and the torsion angle $\chi_{O5B-O5D}$ in  Fig.\ref{fig:NADHdist} that characterizes NA -- AD and NA -- Pyrophosphate (PP) interactions. $E_D$ as function of the distance $R_{C6A-C2N}$ between AD and NA rings is shown in Fig. \ref{fig:dispersion} separately for \emph{cis} and \emph{trans} conformations of NADH in water and methanol. The computed values of $E_D$ are shown in Fig. \ref{fig:dispersion} with black squares that are connected with blue lines for the sake of clarity.

At short and medium distances at $R_{C6A-C2N} < 8$ {\AA} the conformations can be attributed to the folded group where interaction between AD and NA dominates. This effect is known to be due to $\pi$-stacking interaction~\cite{Heiner2017}. At long NA--AD distances when $R_{C6A-C2N} \geq 8$ {\AA} the conformers can be attributed to the unfolded group where the interaction between AD and NA moieties is small, or negligible.

As can be seen in Fig. \ref{fig:dispersion} the energy $E_D$ decreases in general with increasing of the distance between AD and NA rings. However as can be seen in Fig. \ref{fig:dispersion} the dispersion energy $E_D$ contains a pronounced non-monotonic region at the distances $R_{C6A-C2N}$ shorter than about  7~{\AA}. According to the data presented in Tables~\ref{tab:nadh_conf_geom_water} and \ref{tab:nadh_conf_geom_methanol} in \emph{SI}, this non-monotonic behavior correlates with sharp changes of the interatomic distance $R_{C1B-C1D}$ and the torsion angle $\chi_{O5B-O5D}$  shown in Fig.\ref{fig:NADHdist}. The nonmonotonic behavior can be attributed to the interaction between NA and Pyrophosphate (PP) moieties. This conclusion suggests that the widely used consideration of only folded and unfolded conformation groups is insufficient for understanding  the excited state dynamics in NADH because interaction between the NA and PP moieties is sometimes also very important. An important role of  NA -- PP interactions in NADH was discussed recently  by Smith and Tanner \cite{Smith2000}.

The direction of the transition dipole moment (TDM) with respect to the NA plane is an important characteristic for excited state symmetry analysis and for verifying  the validity of the models developed. TDM components under excitation of several NADH conformers from their ground state to the first excited state was computed in aqueous and methanol solutions. The results are presented in \emph{SI} in Tables \ref{tab:TDM_xyz_angle_water} -- \ref{tab:TDM_xyz_angle_water1}.

%%%%%%%%%%%%%%%%%%%%%%%%%%%%%%%%%%%%%%%%%%%%%%%%%%%%%%%%%%%%%%%%%%%%%%%%%%%%%%%%%
\section{V. Discussion}
\label{S:discuss}
\subsection{Excited state lifetime heterogeneity in NADH  }
 \label{sec:heterogeneity}
As was already mentioned in the Introduction, several authors \cite{Gafni1976,Couprie1994a,Ladokhin1995,Visser1981} suggested a model implying a strong relationship between two decay times observed in NADH (see Fig. \ref{fig:times}) and the relative concentration of the folded and unfolded conformations. However this model failed to explain relatively stable values of the preexponential factors in NADH in solutions and existence of two decay times in mononucleotides NMNH ($\beta$-NA mononucleotide) and MNH (1-methylNA) that lack the AD moiety.\cite{Krishnamoorthy1987,Kierdaszuk1996,Blacker2019}

Our experimental data shown in Figs. \ref{fig:ratio} and \ref{fig:Nfol} also do not support this model. As shown in Fig. \ref{fig:ratio} the preexponential factor $a_2$ that refers to the longer lifetime in Table \ref{tbl:1} is practically constant and equal to $a_2=0.24$  at methanol concentrations below 70\% and rises up to 0.4 at higher concentrations. At the same time the relative ratio of the folded conformation $N_{fol}$ is known to  decrease dramatically with methanol concentration and approaches zero value at about 70-80\% MeOH~\cite{Oppenheimer1971,McDonald1972,Hull2001,Heiner2017} (see also Fig.\ref{fig:Nfol} below).

These results  supported the explanation~\cite{Krishnamoorthy1987,Kierdaszuk1996,Blacker2013,Blacker2019} of the lifetimes heterogeneity by inherent photoprocesses occurring in the NA chromophore of NADH. We in general agree with the recent suggestion by Blacker et al.\cite{Blacker2013,Blacker2019} who drew attention on the profound role of \emph{cis} and \emph{trans} configurations of the NA ring for explanation of the lifetime heterogeneity. However as shown below the nature of the particular lifetime values is still controversial and remains a subject of discussions. As reported by Scott et al.\cite{Scott1970} the fluorescence quantum yield of NADH in water is 0.019 which means that the measured fluorescence decay times reflex mostly the non-radiative decay of NADH excited states. Particular routes of this decay in NADH are still unknown. These can be either interactions of excited NADH molecules with surrounding solvent molecules, or intramolecular processes like internal conversion via conical intersections to the ground electronic state, or intersystem crossing to the triplet manifold. \cite{Lakowicz97a}

The dependence of the decay times  on methanol concentration shown in Fig.\ref{fig:times} and on water-methanol solution viscosity in Fig.\ref{fig:rot} do not support the hydrodynamic origin of the non-radiative relaxation decay mechanism suggested by  Blacker et al.\cite{Blacker2013,Blacker2019}. According to the hydrodynamics theory the decay time must be proportional to the solution viscosity: $\tau = k^{-1}_{nr} \sim \eta$, where $k_{nr}$ is a non-radiative decay rate. \cite{Blacker2013} However the decay times $\tau_1$ and $\tau_2$ in Fig.\ref{fig:times} do not follow the change of the solution viscosity as a function of methanol concentration shown in Fig.\ref{fig:rot}.

We believe that the heterogeneity in the measured decay times is due to different charge distributions in the \emph{cis} and \emph{trans} configurations of the NA ring that results in different electrostatic field distributions in these two configurations. The influence of external electric field on the fluorescence decay time was recently investigated by Nakabayashi \emph{et al.} \cite{Nakabayashi2014} who demonstrated that application of an external electric field increases the probability of non-radiative decay in the NADH excited state and therefore shortens the measured fluorescence decay time. The increase of non-radiative decay rate was attributed to the $\pi\pi^*$ character of the electronic transition of the nicotinamide moiety.\cite{Nakabayashi2014} Moreover, Nakabayashi \emph{et al.} \cite{Nakabayashi2014} studied the fluorescence mean decay time in NADH as a function of the solution polarity and reported its significant shortening upon the increase of polarity.

\begin{figure}[h]
\centering
\includegraphics[scale=0.8]{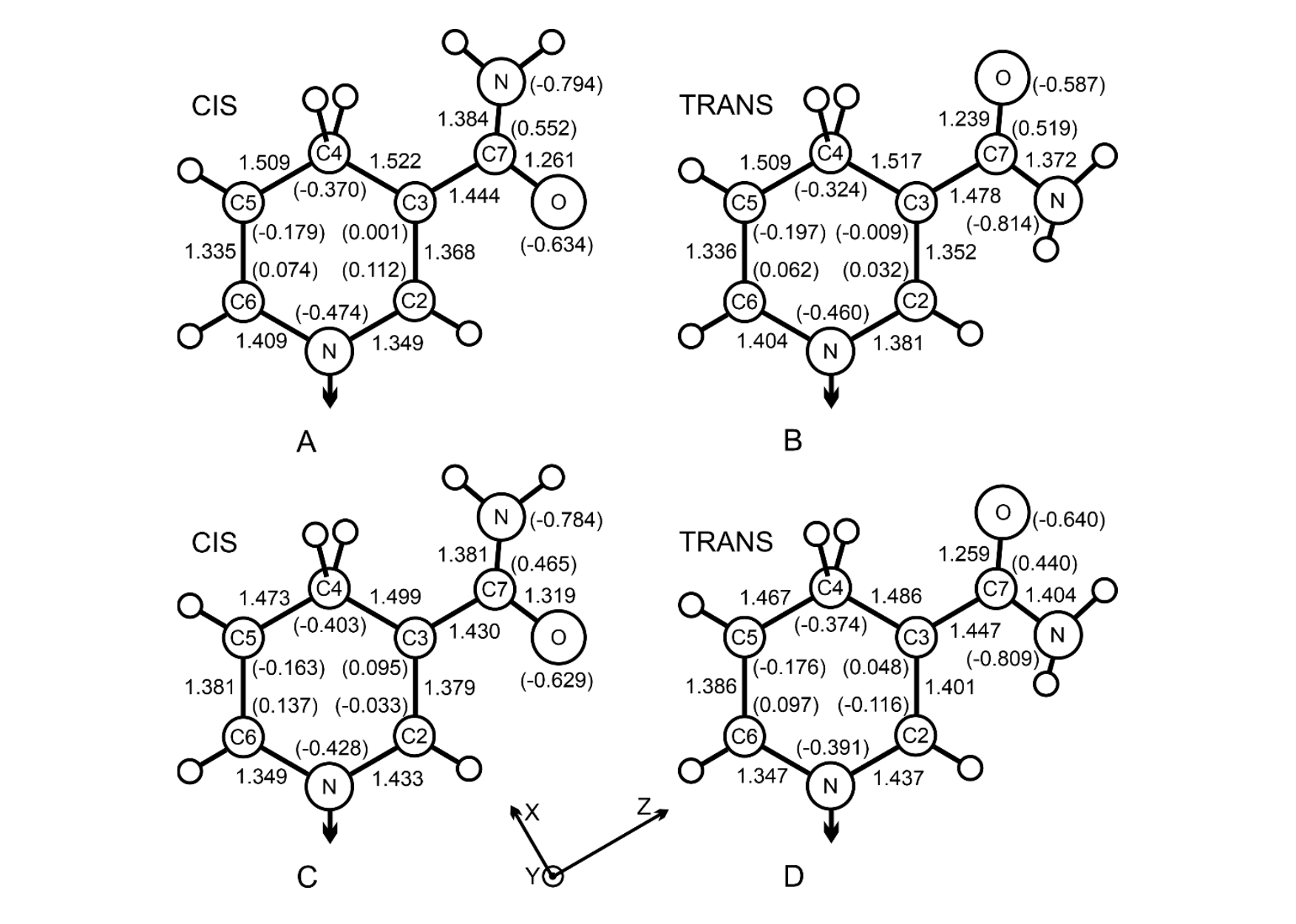}
\caption{Schematic of the NA ring of NADH conformation N11 in water. \\
A and B are $cis$ and $trans$ conformations, respectively, in the ground electronic state. \\
C and D are $cis$ and $trans$ conformations, respectively, in the relaxed first excited electronic state. \\
The coordinate frame  in the figure bottom refers to the TDM components shown in Table~\ref{tab:TDM_xyz_angle_water} in \emph{SI}.}
\label{fig:ring}
\end{figure}

The schematics of the \emph{cis} and \emph{trans} nuclear configurations of the NA ring of NADH conformation N11 (see Tables \ref{tab:nadh_conf_geom_water} and \ref{tab:nadh_conf_geom_water1} in \emph{SI}) in aqueous solution are shown in Fig.\ref{fig:ring}. Earlier Wu \emph{et al}. \cite{Wu1993} reported the results  of \emph{ab initio} calculations of the ground state NADH in vacuum and concluded that the \emph{cis} configuration is more planar and stabilized by electrostatic interaction between the negatively charged amide oxygen atom and the positively charged carbonyl atom C2, while the \emph{trans} configuration is somewhat non-planar and adopts a boat conformation due to the hydrogen bonding between the same oxygen atom and one of the two out-of-plane hydrogen atoms.  These geometry changes characterise the ring puckering effect \cite{Wu1993,Wu1995}.

The results of our calculations of the geometry and charge distributions in the NADH ground and first excited states in aqueous solution presented in Fig.\ref{fig:ring} and in Tables \ref{tab:nadh_conf_geom_water} and \ref{tab:nadh_conf_geom_water1} in \emph{SI} are in general agreement with the conclusions made by Wu \emph{et al} \cite{Wu1993} however contain new important information.  Bonding and dihedral angle values that describe the deformations of the NA moiety in the conformation 11 of NADH in the ground and first excited states in aqueous solution are collected in Tables \ref{table:11BN} and \ref{table:11DH} in \emph{SI}.

In brief, the conclusions on the nuclear geometry and the ring puckering in NA moiety in NADH are as follows:

1. According to \emph{ab initio} calculations the NA moiety is never planar, the amide group in both \emph{cis} and \emph{trans} configurations is out of NA plane. As can be seen in Table \ref{tab:nadh_conf_geom_water} in \emph{SI} in the \emph{trans} configuration the effect is much more pronounced, and the torsion angle $\varphi_{C2N-Nn}$ in some conformations exceeds 30$^\circ$. However, in the first approximation the NA moiety can be treated as planar (see the last subsection of this section), because the deviations from planarity are in general not large.

2. The slope of the C3--C7 bond that is characterized by bonding angles in Table \ref{table:11BN} in \emph{SI} differs in the \emph{cis} and \emph{trans} configurations by several degrees due to the attraction of the negatively charged amide oxygen atom to the carbonyl atom C2 and an out-of-plane hydrogen atom, respectively.

3. The NA ring deformation (puckering) that is characterized by dihedral angles in Table \ref{table:11DH} in \emph{SI} differs from zero in both \emph{cis} and \emph{trans} configurations. The puckering is noticeably more pronounced in the excited state \emph{trans} configurations.

According to the results of our calculations shown in Tables \ref{tab:nadh_ex_energy_water} and \ref{tab:nadh_ex_energy_methanol} in \emph{SI} vertical excitation energies from the ground to the first excited state  do not differ significantly from each other  for the \emph{cis} and \emph{trans} configurations in various  NADH conformations (except in only a few of them). At the same time, as can be seen in Fig.\ref{fig:ring} and confirmed by the torsion angle $\varphi_{C2N-Nn}$  values in Table \ref{tab:nadh_conf_geom_water} in \emph{SI} the charge distributions of \emph{cis} and \emph{trans} configurations differ from each other significantly. These charge distributions result in different electric field configurations for \emph{cis} and \emph{trans} forms that likely leads to the decay time heterogeneity in NADH observed in experiment.

The angle $\gamma$ between the direction of fluorescence TDM in \emph{cis} and \emph{trans} configurations of the conformation 11 and the NA ring plane is  shown in fifth column in Table~\ref{tab:TDM_xyz_angle_water1} in \emph{SI}.  As can be seen in Table~\ref{tab:TDM_xyz_angle_water1} the fluorescence TDM is not parallel to the NA plane. Difference between the angles $\gamma$ related to  the \emph{cis} and \emph{trans}  configurations  exceeds 10$^\circ$ indicating different fluorescence conditions in \emph{cis} and \emph{trans} configurations and supporting the above suggestion.

The internal relaxation mechanism mentioned above that describes the decay time heterogeneity in NADH  is assumed to be the major one as it explains the relative stability of the observed fluorescence lifetimes in different conformers and in various solutions. At the same time a slight but noticeable rise of the decay times $\tau_1$ and $\tau_2$ in Fig.\ref{fig:times} with the increase of methanol concentration can be treated as a minor effect of the slowing of non-radiative relaxation  due to solution polarity decrease.

The preexponential coefficients $a_2$ and $a_1=1-a_2$ in eq.(\ref{eq:iso}) describe the relative population of the the \emph{cis} and \emph{trans} potential energy wells. These populations reflect the balance between the incoming and outgoing population fluxes in each well. The rotational barrier for\emph{ cis} to \emph{trans} interconversion was calculated by Wu \emph{et al}. \cite{Wu1993} and found to be about 7 kcal/mol that is an order of magnitude higher than vibrational energy at room temperature. Our calculations of  NMNH given in the next section also result in comparable rotational barrier heights. The outgoing fluxes is known from experiment being coincide with the decay time values, while sophisticated molecular dynamics calculations are needed for determination of the incoming fluxes.

For qualitative analysis we used a model developed in our recent publication. \cite{Gorbunova20b} According to this model the initial excitation laser pulse promotes the molecule from the ground state to the highly excited vibration energy levels of the electronic excited state. The energy of these vibrational states is higher than the rotational barrier separating the \emph{cis} and \emph{trans} potential energy wells. The following vibrational relaxation occurring in the picosecond time domain leads to the population of the potential wells depending on the details of the molecular structure and of  interactions with solvent molecules. As can be seen in Fig.\ref{fig:ratio} at the methanol concentrations lower than 60\%  the coefficient $a_2$ was a constant, $a_2 \approx 0.25$  within the experimental error bars, and at higher methanol concentrations it grew up to $a_2 \approx 0.4$.

However, the quantitative interpretation of the preexponential coefficient dependence in Fig.~\ref{fig:ratio} is still a  challenging task as it needs several hard and still unresolved problems to be addressed. These are: (i) determination of the ground state population distribution of NADH conformations including their $cis$ and $trans$ configurations in various solutions, (ii) determination of the two-photon excitation probabilities for each of these conformations from the ground electronic state to the high-laying vibrational states of the first electronic excited state, and (iii) population of the excited state $cis$ and $trans$ potential wells during picosecond vibrational relaxation.

\subsection{Lifetime heterogeneity in NMNH}
For clarifying the role of the amide group rotation by the dihedral angle $\varphi$ in NADH in Fig.\ref{fig:NADH}A the theoretical study of a more simple NA-containing molecule NMNH has been performed. Full geometry optimization of a number of NMNH conformers in vacuum in the ground and first excited states has been carried out at the B3LYP-D3BJ/6-31G* level. The schematic of NMNH is presented in Fig.\ref{fig:NADH}B. An important feature of NMNH is that it lacks the AD moiety, however several authors reported the observation of two, or even three fluorescence lifetimes in NMNH in solutions.\cite{Krishnamoorthy1987}.  Potential energy surfaces (PES) of several NMNH conformations were computed for better understanding of the nature of this effect.

\begin{figure}[h]
\centering
\includegraphics[scale=0.7]{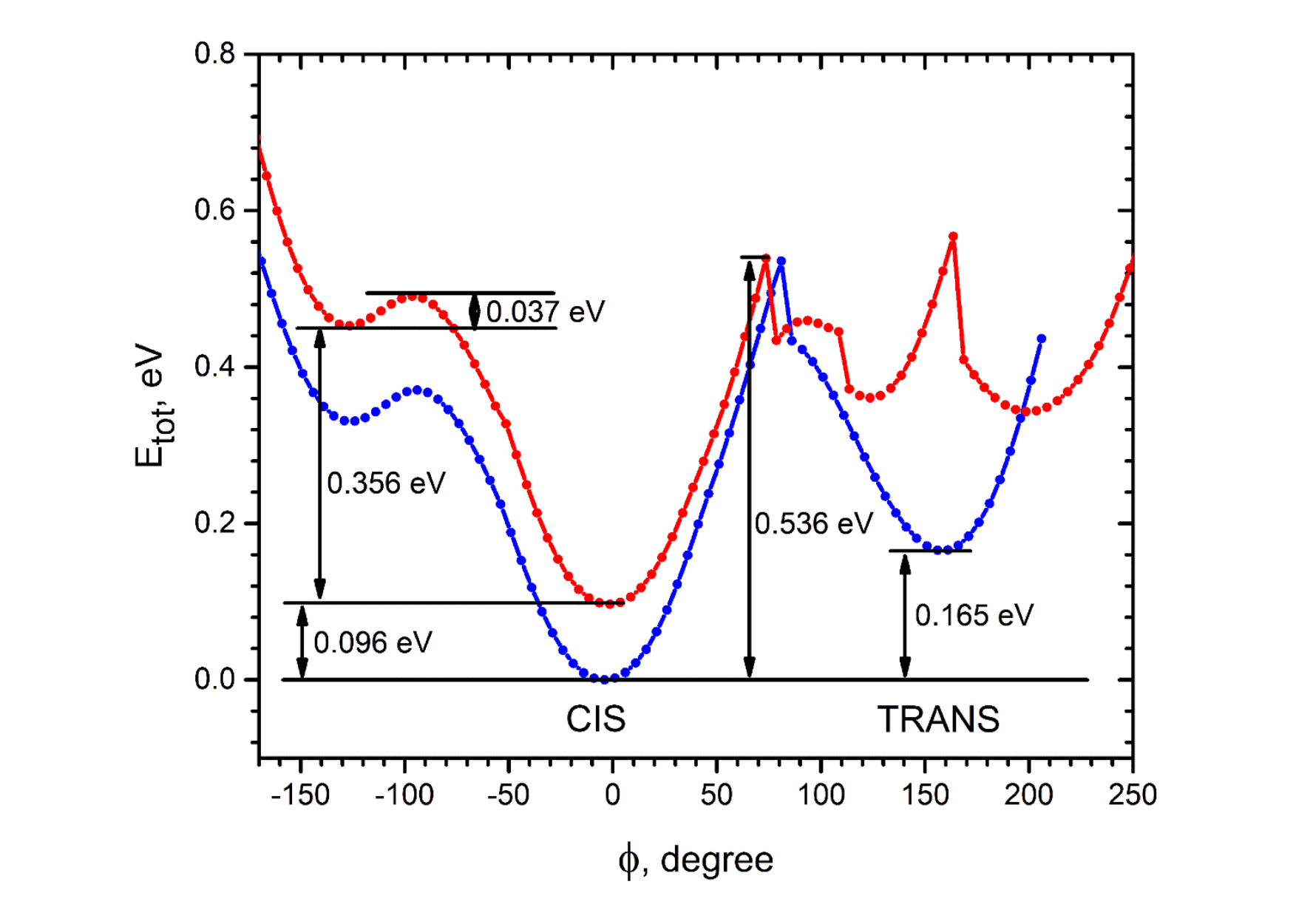}
\caption{Relaxed PES scans of the amide group torsion angle $\phi$ in the NMNH ground electronic state in vacuum. \\
The angle $\phi=\varphi+\pi$, where $\varphi$ is the dihedral angle C2N--C3N--C7N--Nn in Fig.\ref{fig:NADH}B.\\
Blue and red potential curves represent two selected NMNH conformations from many others possible.  }
\label{fig:NMNH}
\end{figure}

The relaxed PES scans of the amide group torsion angle $\phi=\varphi+\pi$ in the NMNH ground electronic state in vacuum are shown in Fig.\ref{fig:NMNH}.

Due to strong hydrogen bonding between the amide oxygen atom and the nearest hydrogen atom from the phosphate group the amide group rotation by the angle $\varphi$ in Fig.\ref{fig:NADH}B is significantly restricted and results in simultaneous change of equilibrium positions of all other nuclei. Therefore  the $cis$ and $trans$ configurations in NADH cannot be treated independently being closely interrelated with other molecular conformations. Moreover, the calculated rotational potential curves depend on the initial value of the torsion angle $\varphi$ and have strong interaction between each other at certain angles $\varphi$. This feature results in a large number of rotational energy curves having one, or several local minima each. Only two of these potential curves are shown in Fig.\ref{fig:NMNH} for simplicity. The red curve in Fig.\ref{fig:NMNH} is a relaxed PES scan of the torsion angle calculated from the potential minima at $\phi=-4^{\circ}$ related to $cis$ molecular geometry. The blue curve in Fig.\ref{fig:NMNH} is a relaxed PES scan calculated from the potential minima at $\phi=156^{\circ}$ related to $trans$ molecular geometry.

As can be seem in Fig.\ref{fig:NMNH} $cis$ and $trans$ molecular geometries in NMNH are characterised by deep minima separated by a relatively high potential barrier. This founding suggests that the double-minimum potential energy curves in Fig.\ref{fig:NMNH} related to $cis$ and $trans$ molecular geometries are responsible for two fluorescence lifetimes observed in NMNH in water \cite{Krishnamoorthy1987}.

 \subsection{Rotational diffusion and the determination of relative conformation concentrations }
 \label{sec:rotation}
 The rotation diffusion time $\tau_{r}$ in Fig.\ref{fig:rot} can be described by the generalized Stokes-Einstein-Debye expression\cite{Dote1981,Anderton1994}
\begin{eqnarray}
\label{eq:SEformula}
\tau_r=fC\frac{\eta V_M}{kT},
\end{eqnarray}
where $k$ and $T$ are the Boltzmann constant and absolute temperature respectively, $V_M$ is a van der Waals solute molecule volume, $\eta$ is the macroscopic solvent viscosity, parameter $f\geq 1$  is a shape factor introduced to account for the non-spherical shape of solute molecules, and the term $C$ signifies the extent of coupling between the solute and solvent.

It is known as the boundary condition parameter \cite{Anderton1994} ($C$ = 1 for "stick", $C$ = 0 for "slip" boundary condition). In general, the factors $f$ and $C$ in eq.(\ref{eq:SEformula}) are strongly interrelated in their effect on solvent-solute friction. The shape factor $f$ is usually taken in the form $f=f_{stick}$ that is a well-specified hydrodynamical frictional coefficient for stick boundary conditions that depends only on the shape of the rotating molecule and can be calculated analytically for any symmetric top molecule \cite{Dote1981}. The boundary condition parameter $C$ depends on relative size of the solute compared to the solvent and is accounted for various effects of solvent-solute interaction: hydrodynamic friction, solute molecule shape, complexes formation, and free volume of the solvent. \cite{Dote1981,Anderton1994}   When the dimensions of a solute molecule are sufficiently large compare with a solvent molecule the parameter $C$ can be presented as $C=f_{slip}/f_{stick}$ \cite{Anderton1994}, where $f_{slip}$ is a hydrodynamical frictional coefficient for slip conditions. The values of $f_{slip}$ related to the  limiting case of pure slip conditions were tabulated for any prolate and oblate top molecule by Hu and Zwanzig \cite{Hu1974}, however very important intermediate cases between the slip and stick boundary conditions can hardly be treated theoretically. When the dimensions of the solute molecules are much larger than the solvent molecules the parameter $C$  usually approaches unity that is related to the stick boundary conditions \cite{Dote1981,Anderton1994}.

Equation (\ref{eq:SEformula})  is known to be valid at relatively low solvent viscosities of approximately $\eta \leq$ 0.1 Pa$\cdot$s.    As can be seen in eq.(\ref{eq:SEformula}) the rotation diffusion time $\tau_r$  is proportional to the solute van der Waals molecular volume $V_M$. The dry volume of NADH was calculated using Edward’s increment method \cite{Edward1970} and found to be 0.488 nm$^3$. A complementary hydration volume $V_{hyd}$ can be added to this volume for taking account of the hydration effect.

The shape factors $f_{stick}$ were calculated for several NADH conformations in Table \ref{tab:nadh_conf_geom_water} in \emph{SI} at the stick boundary conditions using eq.(2) in ref. \cite{Anderton1994} They are presented in Table \ref{table:f}.  The calculation was carried out assuming that NADH can be approximated as a prolate ellipsoid with the ratio of the major and minor axes $R_{max}/R_{min}$.

\begin{table}
  \caption{Axes ratios and the shape factors $f_{stick}$ for several NADH conformations in Table \ref{tab:nadh_conf_geom_water} in \emph{SI}. }
  \label{table:f}
\begin{tabular}{ccc}
\hline
 Conf. No & $R_{max}/R_{min}$ & $f_{stick}$  \\ \hline
1   & 1.19  & 1.06 \\ \hline
10  &  1.68  & 1.29  \\ \hline
11  & 2.18  & 1.63 \\ \hline
12  & 2.96  & 2.30 \\
\hline
\end{tabular}
\end{table}

The conformation No 1 in the first raw in Table \ref{table:f} is the most compact folded one and the conformation No 12 in the fourth raw is the most extended unfolded one while the conformations No 10 and 11 belong to the intermediate group.

In the conditions of our experiments the signals in Fig.\ref{fig2}  were satisfactory approximated by a three-exponential expression in eqs.(\ref{eq:y conv})-(\ref{eq:aniso}) containing two isotropic decay times $\tau_1$ and $\tau_2$ in eq.(\ref{eq:iso}) and one rotational diffusion time $\tau_r$ in eq.(\ref{eq:aniso}). A single-exponential anisotropic distribution in eq.(\ref{eq:aniso}) in fact represented an unresolved multiexponential anisotropic signal containing contributions from many NADH conformations (see Tables \ref{tab:nadh_conf_geom_water} and \ref{tab:nadh_conf_geom_methanol} in $SI$), however the relative populations of these conformations are still unknown. Therefore the interpretation of the dependence of the rotational diffusion time $\tau_r$ on methanol concentration in Fig.\ref{fig:rot} was done by considering a simplified model presenting $\tau_r$ in the form of a sum of contributions from the folded and unfolded conformations (the details are given in  \emph{SI}):
\begin{eqnarray}
\label{eq:expansion}
\frac{1}{\tau_r}\approx\frac{1-(1-N_{fol})\chi(\eta)}{\tau_{fol}}+\frac{N_{un}\chi(\eta)}{\tau_{un}},
\end{eqnarray}
where $\tau_{fol}$ and $\tau_{un}$ are rotational diffusion times that refer to the pure folded and unfolded boundary conditions, respectively, $N_{fol}=N_{fol}(\eta)$ and $N_{un}=N_{un}(\eta)=1-N_{fol}$ are relative ground state NADH concentrations of folded and unfolded conformations in the solution.

The term  $\chi(\eta)$ in eq.(\ref{eq:expansion}) is the quantum yield of laser induced fluorescence from the first excited state of NADH as a function of viscosity $\eta$ in the water-methanol solution. It is defined by the expression:
\begin{eqnarray}
\label{eq:chi}
\chi(\eta)=\frac{I_{fl}(M)}{I_{fl}(\eta)},
\end{eqnarray}
where  $I_{fl}(\eta)$ is the fluorescence intensity as function of $\eta$ and $I_{fl}(M)$ is the fluorescence intensity in pure methanol.

\begin{figure}[h]
\centering
\includegraphics[scale=0.5]{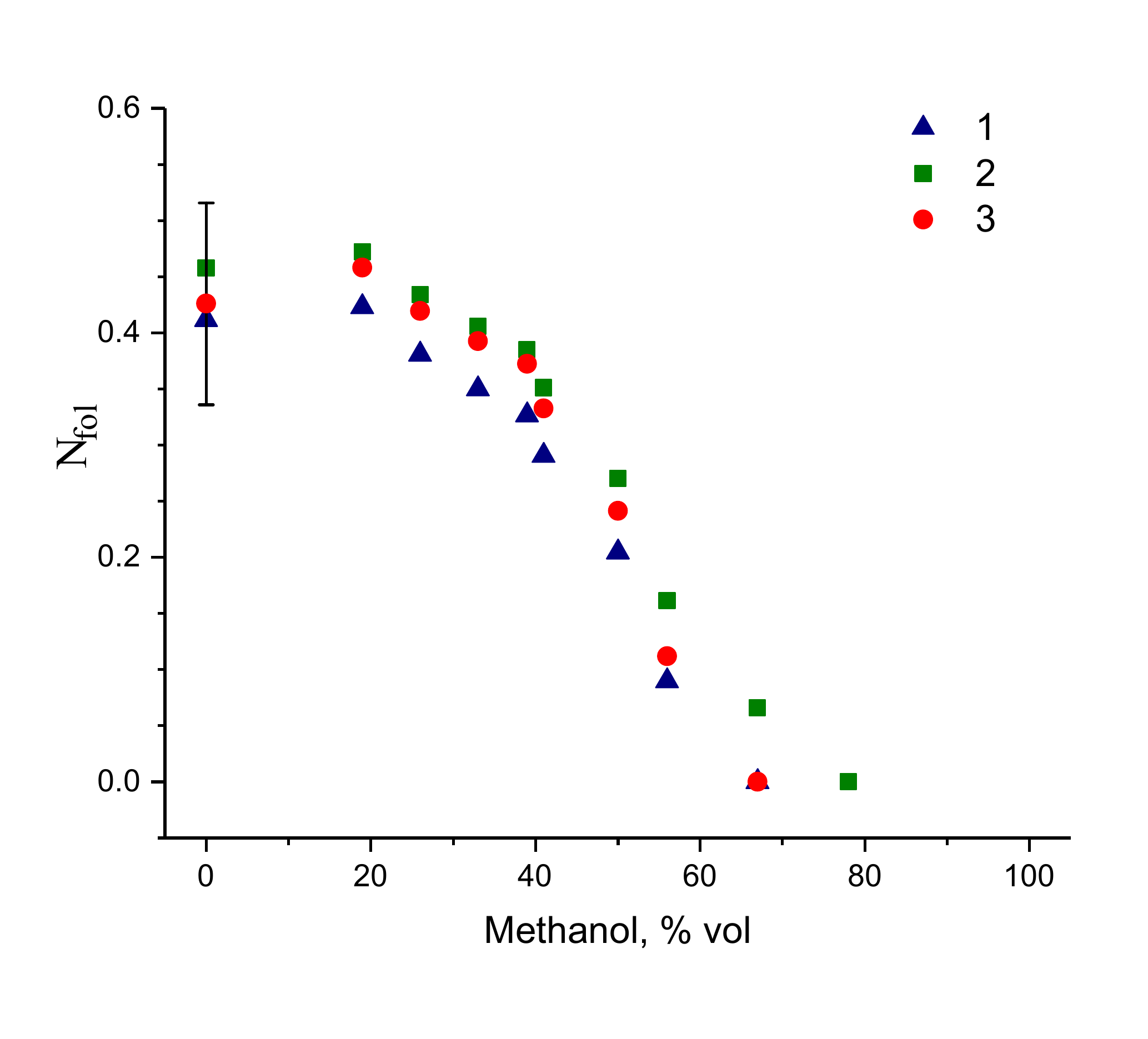}
\caption{Relative folded confirmations $N_{fol}$  as function of the methanol concentration. \\
1. $C_{un}$ = 0.64, $C_{fol}$ = 0.70, $V_{un}$ = 0.528 nm$^3$, $V_{fol}$ = 0.548 nm$^3$ \\
2. $C_{un}$ = 0.75, $C_{fol}$ = 0.80, $V_{un}$ = 0.488 nm$^3$, $V_{fol}$ = 0.488 nm$^3$ \\
3. $C_{un}$ = 0.70, $C_{fol}$ = 0.85, $V_{un}$ = 0.488 nm$^3$, $V_{fol}$ = 0.488 nm$^3$}
\label{fig:Nfol}
\end{figure}

As known~\cite{Freed1967,Heiner2017} $\chi(\eta)$ decreases monotonously from about 1.6 to 1 with increase of methanol concentration. For analysis, the rotational diffusion times $\tau_{fol}$ and $\tau_{un}$ in eq.(\ref{eq:expansion}) were calculated from the Stokes-Einstein-Debye formula in eq.~(\ref{eq:SEformula}), where  the shape factors $f_{stick}(fol)$ and $f_{stick}(un)$ were taken from the first and fourth rows of Table~\ref{table:f}, respectively, and the values $C_{fol}$, $C_{un}$, $V_{fol}$, $V_{un}$ were used as fitting parameters.  The concentration $N_{fol}$ determined from eq.(\ref{eq:expansion}) using the values of $\tau_r$ and $\eta$ in Fig.\ref{fig:rot} and the values of $\chi(\eta)$ from ref.\cite{Freed1967} at several  sets of the parameters $C_{fol}$, $C_{un}$, $V_{fol}$, and $V_{un}$ that are relevant for our experimental conditions and consistent with the results of previous studies are given in Fig.\ref{fig:Nfol}.

As can be seen in Fig.\ref{fig:Nfol} the relative concentration of the folded conformation $N_{fol}$ in NADH  in pure water was about 0.4  and then decreased down to zero with increasing methanol concentration due to denaturation.  This conclusion agrees qualitatively  with the results of all other early studies.\cite{Freed1967,Hull2001,Heiner2017} The distributions of conformations in NADH in water and water-methanol solutions were earlier studied on the basis of the two-states folded-unfolded model by several groups  by means of various experimental methods: NMR \cite{Oppenheimer1971,McDonald1972}, energy transfer \cite{Hull2001,Heiner2017}, and fluorescence lifetime \cite{Couprie1994a} measurements. The advantage of the method used in this paper is that it is based on the study of the anisotropy  of excited molecular axes distribution and refers directly to a solute molecule shape.  The method does not need a sophisticated model for data extraction from experiment as NMR measurements do. Moreover it allows to avoid a typical problem of optical methods dealing with a necessity to account for non-radiative transitions.

$N_{fol}$ value in pure water at 20$^\circ$ shown in Fig.\ref{fig:Nfol} was found to be about $0.4\pm 0.1$ at all values of the fitting parameters used. This value agrees within experimental error bars with the results reported by Oppenheimer et al. \cite{Oppenheimer1971}, $(0.36)$ and by Heiner et al., $(0.26\pm 0.06)$. However, our result is somehow larger than that reported by McDonald \cite{McDonald1972} (0.24) and smaller than that reported by Hull et al. \cite{Hull2001} (0.55).

Note that the quantity $N_f$ determined from NMR and energy transfer experiments  refs.\cite{Oppenheimer1971,McDonald1972,Hull2001,Heiner2017} is not exactly the same as the quantity $N_{fol}$ determined in this paper. The former deals mostly with the ratio of stacked and unstacked conformations and refers to the extent of chemical interaction between NA and AD rings, while the latter refers mostly to the geometrical properties of NADH conformations. Therefore, some disagreement between the results obtained by the former and the latter methods is possible and should not surprise.

 \subsection{Determination of the components of the two-photon excitation tensor $\mathbf{S}$. }
 \label{sec:symmetry}
As can be seen in Figs.~\ref{fig:anisotropy}, \ref{fig:Omega}, and in Table~\ref{tbl:1} the excitation anisotropy parameters $r_l$, $r_c$, and the parameter $\Omega$ determined from experiments were found to be almost independent of methanol concentration. Therefore, we assumed that these parameters did not depend on the solution composition and used for analysis their mean values calculated from Table \ref{tbl:1} and denoted as $\bar{r}_l$ $\bar{r}_c$, and $\bar{\Omega}$. As can be seen in Table \ref{tbl:1} the parameter mean values are equal to $\bar{r}_l=0.47$ $\bar{r}_c=-0.23$, and $\bar{\Omega}=0.8$, ane the relationship between the anisotropy parameters follows almost perfectly the formula $r_l = -2r_c$. The same relationship was also found in our recent study of two-photon excited fluorescence in indol \cite{Sasin18}. According to the model developed in our paper  \cite{Sasin18} different signs and values of the anisotropies  $r_l$ and $r_c$ were due to different types of alignment produced by linearly polarized (LP) and circularly polarized (CP) photons in the molecular axes distribution. In fact, in the case of one-photon excitation the formula $r_l = -2r_c$ is exact and should be valid for any molecules. However for two-photon excitation the formula implies a certain relationship between the components of the two-photon excitation tensor $\mathbf{S}$.

The determination of the components of the  two-photon excitation tensor $\mathbf{S}$ from the experimental data was carried out on the basis of eqs.~(\ref{eq:MK}), (\ref{eq:SR}). The mean values of the fluorescence parameters $\bar{r}_l$, $\bar{r}_c$, $\bar{\Omega}$ determined from experiment and the normalized molecular parameters $ M_K({R,R'})$ calculated according to  eqs.~(\ref{eq:rl})--(\ref{eq:Omega}) are collected in Table~\ref{table:two1}.

 \begin{table}
  \caption{Molecular parameters determined from experiment. }
  \label{table:two1}
\begin{tabular}{cccccc}
\hline
 $\bar{r}_l$ & $\bar{r}_c$ & $\bar{\Omega}$ &\large$\frac{M_0(22)}{M_0(00)}$ & \large$\frac{M_2(02)}{M_0(00)}$ & \large$\frac{M_2(22)}{M_0(00)}$ \\ \hline
0.47 & -0.23 & 0.8   & 1.28$\pm$0.06  & 1.86$\pm$0.08 & 1.72$\pm$0.09 \\
\hline
\end{tabular}
\end{table}

As can be seen in eq.~(\ref{eq:MK}) the molecular parameters $M_K(R,R')$ depend not only on the   two-photon excitation spherical tensor components $S_{2\gamma}$ in eq.~(\ref{eq:SR}), but also on the fluorescence TDM components $F_{1q_{fl}}$ in eq.~(\ref{eq:Fq}). However, by choosing a \emph{fluorescence} body frame, where axis Z$_{fl}$ is parallel to the fluorescence transition dipole moment direction two from three fluorescence TDM components $F_{1\pm 1}$ in eq.~(\ref{eq:MK}) were canceled out and only one component $F_{10}=F_z$ remained non-zero. In this case all M-parameters in eq.~(\ref{eq:MK}) were proportional to the term $|F_z|^2$ that could be taken outside from the sum and the particular $M$-parameter expressions could be written in term of the two-photon excitation tensor components in the form of eqs.~(17)--(21) in ref.~\cite{Denicke10}.

The inspection of the calculation data in Tables \ref{tab:TDM_xyz_angle_water}--\ref{tab:TDM_xyz_angle_water1} in \emph{SI} reveals that Z-component of TDM in NADH is an order of magnitude, or more larger than X and Y components. Moreover, as shown in Table~\ref{tab:TDM_xyz_angle_water1} the direction of the TDM calculated in the relaxed excited state molecular frame is almost parallel to the direction of the one-photon TDM calculated in the ground state molecular frame. One can conclude that both molecular frames in Tables \ref{tab:TDM_xyz_angle_water}--\ref{tab:TDM_xyz_angle_water1} are close to the \emph{fluorescence} TDM frame described above.   Therefore, within these subsection the fluorescence TDM in NADH was assumed to be approximately parallel to the NA ring. This frame is drawn in the bottom of Fig.~\ref{fig:ring}.

The Cartesian components of the two-photon excitation tensor $\mathbf{S}$ calculated from the $M$-parameters values in Table~\ref{table:two1} by solving the systems of equations (17)--(21) and (22)--(25) in our paper~\cite{Denicke10} are shown in Table~\ref{table:two2}.

 \begin{table}
  \caption{Two-photon excitation parameters. }
  \label{table:two2}
\begin{tabular}{cccccccccc}
\hline
  \large$\frac{S_{zz}}{Tr \mathbf{S}}$ & \large$\frac{S_{xx}+S_{yy}}{Tr \mathbf{S}}$ & \large$\frac{S_{xx}-S_{yy}}{Tr \mathbf{S}}$ & \large$\frac{|S^2_{xz}+S^2_{yz}|^{1/2}}{Tr \mathbf{S}}$\\ \hline
 0.95$\pm$0.05 & 0.05$\pm$0.05 & -0.0$\pm$0.04 & 0.44$\pm$0.11\\
\hline
\end{tabular}
\end{table}

The normalization factor $Tr \mathbf{S}$ in Table~\ref{table:two2} is equal to $ Tr \mathbf{S} = S_{xx} +S_{yy}+ S_{zz}$.
As can be seen in Table~\ref{table:two2} the longitudinal diagonal Cartesian component $S_{zz}$ of the two-photon excitation tensor dominates over the transversal diagonal components $S_{xx}+S_{yy}$ indicating that the two-photon excitation is parallel to the direction of the fluorescence TDM. As can be shown from eqs. (22)--(25) in ref.~\cite{Denicke10} this component is described only by the normalized value of the molecular parameter $M_2(0,2)$ according to the expression:
 \begin{eqnarray}
\label{eq:Szz}
\frac{S_{zz}}{Tr\mathbf{ S}}=\frac{1}{3} \left[1+\frac{M_2(0,2)}{M_0(00)}\right].
\end{eqnarray}

As can also be seen in Table~\ref{table:two2} another excitation channel described by the off-diagonal  two-photon tensor components $S_{xz}$ and $S_{yz}$ contributed to the experimental signals along with the diagonal two-photon excitation channel mentioned above.  These off-diagonal tensor components have  in general complex values and only the combination $|S^2_{xz}+S^2_{yz}|$ could be determined from our experiment.

Note that the two-photon  tensor components considered in this section describe the molecular excitation by an ultrashort laser pulse at the time $t=0$. In general they contain contributions from many NADH conformations shown in Tables~\ref{tab:nadh_conf_geom_water} and \ref{tab:nadh_conf_geom_methanol} in \emph{SI}. The separation of these contributions within a given experimental accuracy is a challenging problem.

\section{Conclusions}
The decay of polarized fluorescence in NADH dissolved in water-methanol solutions under two-photon excitation at 720~nm by femtosecond laser pulses was studied experimentally and theoretically. A number of fluorescence parameters have been determined from experiment using the global fit procedure and then compared with the results reported by other authors. Interpretation of the experimental results obtained were supported by intensive \emph{ab initio} calculations of the structure of NADH and NMNH in various solutions. The analysis of the results obtained has led to a new explanation of the heterogeneity in the measured decay times in NADH and in NMNH. The explanation ia based on the influence of the internal molecular electric field on the non-radiative decay rates of the excited molecular states. We suggest that different charge distributions in the \emph{cis} and \emph{trans} configurations of the nicotinamide ring result in different electrostatic field distributions that lead to the decay time heterogeneity. A slight but noticeable rise of the decay times $\tau_1$ and $\tau_2$  with methanol concentration was observed that was treated as a minor effect of non-radiative relaxation slowing due to solution polarity decrease. As shown, the consideration of only folded and unfolded conformation groups is insufficient for understanding the excited state dynamics in NADH because interaction between the NA and phosphate moieties were sometimes also very important.  Determination of the rotational diffusion time in water-methanol solutions as a function of methanol concentration allowed for determination of relative concentrations of the folded and unfolded NADH conformations. Two-photon excitation tensor components have been determined from the experimental data. The analysis of the results obtained suggests the existence of two excitation channels with comparable intensities: the diagonal longitudinal channel that dominated by the two-photon tensor component $S_{zz}$, where axis Z is in NA ring plane and off-diagonal channel dominated by the tensor components $|S^2_{xz}+S^2_{yz}|^{1/2}$.

\section{Conflicts of interest}
There are no conflicts to declare.

\begin{acknowledgement}

IAG, MES, and OSV are grateful to Russian Foundation for Basic Research for financial support under the grant No 18-53-34001.
JRS is grateful to the Programa Nacional de Ciencias Básicas de Cuba PNCB: P223LH001-108 (JRS).  IAG and OSV are grateful to BASIS Foundation for financial support under
the grant no. 19-1-1-13-1. The authors are grateful to the Ioffe Institute for providing the equipment used in the experiments.

\end{acknowledgement}

%%%%%%%%%%%%%%%%%%%%%%%%%%%%%%%%%%%%%%%%%%%%%%%%%%%%%%%%%%%%%%%%%%%%%
%% The same is true for SI, which should use the
\begin{suppinfo}
\label{sec:errors}

\section*{Fitting protocol}
The experimental data was processed by a global analysis procedure \cite{Lakowicz97a} using two alternative fitting methods: differential evolution and nonlinear least squares realized in Matlab and in Python3, respectively.  The experimental response function $IRF(t)$ and the relative sensitivity $G$ were experimentally determined initially and then used in the fitting procedure. The value of the relative sensitivity $G$ was measured to be $G=$0.98. Experimental data related to $x$ and $y$ fluorescence decay polarization components were fitted using eqs.(\ref{eq:y conv}) and (\ref{eq:x conv}) separately for linear and circular polarized excitation to determine the global fluorescence parameters $I_l$, $r_l$, $\tau_1$, $\tau_2$, $a_1$, $a_2$, $\tau_r$ and $I_c$, $r_c$, $\tau_1$, $\tau_2$, $a_1$, $a_2$,  $\tau_r$, respectively.   The cost function used in the fitting procedure,  was the total sum of squared deviations for $x$ and $y$ data sets \cite{Lakowicz97a}:
\begin{equation}
\label{eq:cost}
cost = \sum_{i=1}^n \frac{1}{\sigma_y(t_i)}(I_{YYy}(t_i) - I_{YYy}^{Fit}(t_i))^2 + \sum_{i=1}^n \frac{1}{\sigma_x(t_i)}(I_{YYx}(t_i) - I_{YYx}^{Fit}(t_i))^2,
\end{equation}
where $\sigma_y$ and $\sigma_x$ are weighting functions.

The minimization of the cost function in eq. (\ref{eq:cost}) was carried out by constrained optimization using differential
evolution  implemented in \textit{scipy} \cite{Qin2009}.

\section*{General expression for fluorescence intensity under two-photon excitation}
\label{sec:fluores}
A convenient expression for the excited fluorescence intensity $I(t)$ after two-photon excitation in the case of heterogeneity
in the measured lifetimes can be presented in the form: \cite{Shternin10,Denicke10}
\begin{eqnarray}
\label{eq:intens_final}
I(t)&=&-C \,w(t)
\sum_{K_e}\sum_{K_1, K_2}\sum_{R, R'}\left(\left[{
{\mathbf{E}}}_{K_1}(\mathbf{e}_1)\otimes {
{\mathbf{E}}}_{K_2}(\mathbf{e}_2)\right]_{K_e} \cdot { {\mathbf{E}}}_{K_e}(\mathbf{e}_{fl})\right)
\nonumber\\
&\times& \frac{\sqrt{(2K_1+1)(2K_2+1)(2R+1)(2R'+1)}}{2K_e+1}
\left\{\begin{array}{ccc}
K_1 & K_2 & K_e \\
1 & 1 & R' \\
1 & 1 & R
\end{array}\right\}
\nonumber\\
&\times& M_{K_e}(R,R',t),
\end{eqnarray}
where $\textbf{E}_{K_1Q_1}(\mathbf{e}_1)$, $\textbf{E}_{K_2Q_2}(\mathbf{e}_2)$, and $\textbf{E}_{K_eQ_e}(\mathbf{e}_{fl})$ are the light polarization matrices \cite{Zare88b} of the first, second, and
the fluorescence photons, respectively, and $\mathbf{e}_1$, $\mathbf{e}_2$, and $\mathbf{e}_{fl}$ are corresponding photon polarization vectors.

In the conditions of one-color excitation used in our experiments the first and the second excitation photons were the same, therefore $\mathbf{e}_1=\mathbf{e}_2$. Summation in eq.~(\ref{eq:intens_final}) is proceeded over the ranks $K_1$, $K_2$, $K_e$, $R$, $R'$ each in general is limited to the values 0 and 2. The term $w(t)=\left(a_1e^{-\frac{t-t'}{\tau_1}}+a_2e^{-\frac{t-t'}{\tau_2}}\right)$ describes the heterogeneous excited state
decay with the fluorescence lifetimes $\tau_1$ and $\tau_2$. The term $C$ is a normalization constant that is proportional to the square of the excitation light intensity, the term in braces is a 9$j$-symbol. The term in
parentheses eq.(\ref{eq:intens_final}) denotes an irreducible tensor product defined as: \cite{Zare88b}
\begin{eqnarray}
\label{eq:tensor_product}
&&
\left(\left[{
{\mathbf{E}}}_{K_1}(\mathbf{e}_1)\otimes {
{\mathbf{E}}}_{K_2}(\mathbf{e}_2)\right]_{K_e} \cdot { {\mathbf{E}}}_{K_e}(\mathbf{e}_{fl})\right)\nonumber \\
 &=& \sum_{Q_1,Q_2,Q_e}
(-1)^{Q_e}C^{K_eQ_e}_{K_1 Q_1\:K_2 Q_2}\mathbf{E}_{K_1Q_1}(\mathbf{e}_1)
\mathbf{E}_{K_2Q_2}(\mathbf{e}_2) { {\mathbf{E}}}_{K_e-Q_e}(\mathbf{e}_{fl}),
\end{eqnarray}
where $Q_1$, $Q_2$, and $Q_e$ are the components of the ranks $K_1$, $K_2$, and $K_e$ on the laboratory axis Z,
respectively and the term $C^{K_eQ_e}_{K_1 Q_1\:K_2 Q_2}$ is a Clebsch-Gordan coefficient.\cite{Zare88b}

\begin{equation}
\label{eq:D} {\cal D}^{K_e}_{q_e,q'_e}(t) = \sum_k d^{K_e}_{q_e
k}\,e^{-E_{K_e k}t}\,\left(d^{K_e}_{q'_e k}\right)^*.
\end{equation}

The terms $d^{K_e}_{q_e k}$ in eq.~(\ref{eq:D}) are expansion coefficients of the rotational diffusion operator
eigenfunctions over the Wigner $D$-functions and $E_{K_e k}=1/\tau^{(k)}_r$ are diffusion operator eigenvalues. The
index $k$ is limited to the values $k=1 \dots (2K_e+1)$ and ${\cal D}^0_{00}(t)\equiv 1$.

\section*{Ab initio calculations}
The optimized stable ground state geometries of 24 NADH conformations including \emph{cis} and \emph{trans} in water and methanol are shown in Tables~\ref{tab:nadh_conf_geom_water} and \ref{tab:nadh_conf_geom_methanol}, respectively. The optimized geometries of the first excited state  in aqueous solution are presented in Tab.~\ref{tab:nadh_conf_geom_water1}.

\newpage
\begin{table}
\caption{Optimized stable geometries of NADH conformations in their ground electronic state in aqueous solution.\\
The superscript index "+" indicates that the calculation was performed at the B3LYP-D3BJ/6-31+G* level. }
\label{tab:nadh_conf_geom_water}
\begin{tabular}{c c c c c c}
\hline
Conformation  & $R_{C1B-C1D}$, & $R_{C6A-C2N}$, & $R_{Nn-Op}$, & $\chi_{O5B-O5D}$, & $\varphi_{C2N-Nn}$, \\
           & {\AA} & {\AA} & {\AA} & degree & degree\\
\hline
1 $trans$ & $4.993$ & $3.896$ & $3.429$ & $26.185$ & $-14.644$\\
1 $cis$   & $5.026$ & $3.912$ & $5.549$ & $29.582$ & $-175.675$ \\
2 $trans$ & $4.579$ & $4.856$ & $8.458$ & $-22.497$ & $31.639$\\
2 $cis$   & $4.413$ & $3.762$ & $10.319$ & $-24.307$ & $170.134$ \\
3 $trans$ & $6.374$ & $4.275$ & $8.345$ & $-28.231$ & $32.536$\\
3 $cis$   & $6.639$ & $3.420$ & $8.410$ & $-42.308$ & $170.016$ \\
4 $trans$ & $6.407$ & $4.961$ & $3.297$ & $9.158$ & $-15.899$\\
4 $cis$   & $6.421$ & $4.830$ & $5.366$ & $-0.168$ & $-176.19$ \\
5 $trans$ & $4.129$ & $5.927$ & $5.447$ & $25.614$ & $40.384$\\
5 $cis$   & $4.112$ & $6.528$ & $6.454$ & $29.928$ & $-174.831$ \\
6 $trans$ & $7.853$ & $6.043$ & $8.698$ & $62.596$ & $-17.281$\\
6 $cis$   & $7.851$ & $6.169$ & $9.010$ & $64.629$ & $-173.655$ \\
7 $trans$ & $3.966$ & $6.436$ & $3.216$ & $29.946$ & $-10.228$\\
7 $cis$   & $4.007$ & $6.185$ & $5.247$ & $31.159$ & $-175.704$ \\
8 $trans$ & $5.872$ & $6.952$ & $6.861$ & $2.006$ & $21.563$ ()\\
8 $cis$   & $5.645$ & $6.328$ & $6.644$ & $5.043$ & $169.417$ \\
8 $trans^{+}$ & $5.929$ & $7.044$ & $6.924$ & $1.122$ & $24.035$\\
9 $trans$ & $3.809$ & $7.748$ & $9.256$ & $23.264$ & $21.545$ \\
9 $cis$   & $4.755$ & $6.389$ & $9.882$ & $6.469$ & $-176.073$ \\
10 $trans$ & $6.488$ & $11.989$ & $3.430$ & $21.566$ & $-9.077$\\
10 $cis$   & $6.092$ & $10.698$ & $5.379$ & $10.520$ & $-172.019$ \\
11 $trans$ & $10.486$ & $11.557$ & $4.632$ & $145.964$ & $-17.806$\\
11 $cis$   & $7.751$ & $10.655$ & $5.346$ & $-66.713$ & $177.359$ \\
12 $trans$ & $9.771$ & $13.001$ & $8.499$ & $-170.573$ & $-13.732$\\
12 $cis$   & $9.767$ & $13.781$ & $8.372$ & $-169.960$ & $-178.48366 $ \\
\hline
\end{tabular}
\end{table}
\newpage

\begin{table}
\caption{Optimized stable geometries of NADH conformations in their ground electronic state in methanol solution. }
\label{tab:nadh_conf_geom_methanol}
\begin{tabular}{c c c c c c}
\hline
Conformation  & $R_{C1B-C1D},$ & $R_{C6A-C2N},$ & $R_{Nn-Op}$, & $\chi_{O5B-O5D}$, & $\varphi_{C2N-Nn}$,\\
           & {\AA} & {\AA} & {\AA} & degree & degree\\
\hline
1 $trans$ & $4.993$ & $3.895$ & $3.421$ & $26.163$ & $-14.595$\\
1 $cis$   & $5.026$ & $3.907$ & $5.544$ & $29.666$ & $-175.509$\\
2 $trans$ & $4.579$ & $4.859$ & $8.482$ & $-22.801$ & $31.776$\\
2 $cis$   & $4.416$ & $3.759$ & $10.333$ & $-24.576$ & $170.414$\\
3 $trans$ & $6.375$ & $4.268$ & $8.341$ & $-28.661$ & $32.646$\\
3 $cis$   & $6.636$ & $3.419$ & $8.399$ & $-42.439$ & $170.033$\\
4 $trans$ & $6.409$ & $4.967$ & $3.295$ & $9.281$ & $-15.804$\\
4 $cis$   & $6.420$ & $4.837$ & $5.365$ & $-0.514$ & $-175.827$\\
5 $trans$ & $4.058$ & $6.058$ & $5.455$ & $24.385$ & $40.649$\\
5 $cis$   & $4.111$ & $6.527$ & $6.459$ & $29.927$ & $-175.058$\\
6 $trans$ & $7.833$ & $6.044$ & $8.699$ & $62.035$ & $-17.163$\\
6 $cis$   & $7.844$ & $6.169$ & $9.011$ & $64.580$ & $-173.678$\\
7 $trans$ & $3.953$ & $6.441$ & $3.215$ & $30.408$ & $-10.419$\\
7 $cis$   & $4.002$ & $6.186$ & $5.256$ & $31.201$ & $-175.749$\\
8 $trans$ & $5.870$ & $7.644$ & $6.884$ & $2.009$ & $21.979$\\
8 $cis$   & $5.645$ & $6.330$ & $6.640$ & $5.036$ & $169.664$\\
9 $trans$ & $3.800$ & $7.749$ & $9.269$ & $23.113$ & $21.578$\\
9 $cis$   & $4.737$ & $6.198$ & $9.678$ & $7.544$ & $-177.801$\\
10 $trans$ & $6.490$ & $11.984$ & $3.435$ & $20.906$ & $-8.969$\\
10 $cis$   & $6.093$ & $10.686$ & $5.389$ & $9.937$ & $-172.075$\\
11 $trans$ & $10.409$ & $11.824$ & $4.652$ & $137.844$ & $-17.314$\\
11 $cis$   & $7.736$ & $10.645$ & $3.296$ & $-66.383$ & $177.546$\\
12 $trans$ & $9.772$ & $13.799$ & $8.50$ & $-170.764$ & $-13.964$\\
12 $cis$   & $9.768$ & $13.780$ & $8.371$ & $-170.105$ & $-178.437$\\
\hline
\end{tabular}
\end{table}
\newpage
\begin{table}
\caption{Optimized stable geometries of NADH conformations in their  first excited electronic state {relaxed} in aqueous solution.}
\label{tab:nadh_conf_geom_water1}
\begin{tabular}{c c c c c c}
\hline
Conformation  & $R_{C1B-C1D}$, & $R_{C6A-C2N}$, & $R_{Nn-Op}$, & $\chi_{O5B-O5D}$, & $\varphi_{C2N-Nn}$, \\
           & {\AA} & {\AA} & {\AA} & degree & degree\\
\hline
10 $trans$ & $6.721$ & $12.269$ & $3.476$ & $29.041$ & $-10.629$ \\
11 $trans$ & $10.923$ & $11.465$ & $5.289$ & $130.550$ & $-2.074$ \\
11 $cis$ &  $8.832$ & $11.996$ & $5.553$ & $-65.382$ & $171.855$  \\
\hline
\end{tabular}
\end{table}

The interatomic distances $R_{C6A-C2N}$, $R_{C1B-C1D}$, $R_{Nn-Op}$ and torsion angles  $\chi_{O5B-O5D}$ and $\varphi_{C2N-Nn}$ collected in Tables~\ref{tab:nadh_conf_geom_water}--\ref{tab:nadh_conf_geom_water1} are clarified in Fig.~\ref{fig:NADHdist}. $R_{C6A-C2N}$ is a distance between the carbon atoms $C6A$ and $C2N$ that belong to AD and NA, respectively, $R_{C1B-C1D}$ is a distance between  $C1D$ and $C1B$ carbon atoms in two ribose moieties,  $R_{Nn-Op}$ is a distance between the nitrogen atom $Nn$ in the NA group and the oxygen atom $Op$ in the pyrophosphate group. The torsion angle $\varphi_{C2N-Nn}$ describes orientation of the amide group with respect to the NA ring and the torsion angle $\chi_{O5B-O5D}$ ($C1B-O5B-O5D-C1D$), describes the relative orientation of the ribose rings.

Vertical excitation energies of four lowest electron excited states, corresponding one-photon oscillator strengths, and ground state dispersion energies  for twenty four NADH conformers including $\textit{cis}$ and $\textit{trans}$ modifications are presented in Tables~\ref{tab:nadh_ex_energy_water} and \ref{tab:nadh_ex_energy_methanol} in water and methanol solutions, respectively.

\newpage

\begin{table}
\caption{Vertical excitation energies of four lowest excited states, corresponding oscillator strengths ($Osc. str.$) from the ground electronic state, and Grimme-D3(BJ) ground state dispersion energies in $\textit{trans}$ and $\textit{cis}$ NADH conformers in aqueous solution.}
\label{tab:nadh_ex_energy_water}
\begin{tabular}{c c c c c c}
\hline
  & Energy, eV    & Energy, eV.   & Energy, eV    & Energy, eV    & Dispersion\\
Conformation    & $(Osc. str.)$ & $(Osc. str.)$ & $(Osc. str.)$ & $(Osc. str.)$ & energy,\\
  & Ex. state 1       & Ex. state 2       & Ex. state 3       & Ex. state 4       & Hartree, 10$^{-4}$\\
\hline
1 $trans$ & $3.12(0.161)$ & $3.67(0.019)$ & $3.95(0.019)$ & $4.22(0.029)$ & $2023$\\
1 $cis$   & $3.27(0.207)$ & $3.75(0.000)$ & $3.97(0.012)$ & $4.34(0.008)$ & $2002$\\
2 $trans$ & $3.57(0.154)$ & $3.74(0.012)$ & $4.36(0.007)$ & $4.41(0.007)$ & $1886$\\
2 $cis$   & $3.39(0.121)$ & $3.70(0.047)$ & $4.28(0.021)$ & $4.31(0.024)$ & $1921$\\
3 $trans$ & $3.45(0.153)$ & $3.50(0.021)$ & $4.23(0.001)$ & $4.52(0.043)$ & $1964$\\
3 $cis$   & $3.23(0.040)$ & $3.52(0.151)$ & $4.00(0.002)$ & $4.41(0.188)$ & $1970$\\
4 $trans$ & $3.14(0.151)$ & $3.42(0.034)$ & $4.10(0.004)$ & $4.33(0.066)$ & $1964$\\
4 $cis$   & $3.18(0.165)$ & $3.50(0.039)$ & $4.16(0.002)$ & $4.27(0.055)$ & $1956$\\
5 $trans$ & $3.43(0.119)$ & $3.54(0.038)$ & $4.16(0.005)$ & $4.62(0.333)$ & $2034$\\
5 $cis$   & $3.57(0.121)$ & $3.66(0.058)$ & $4.29(0.002)$ & $4.53(0.001)$ & $1971$\\
6 $trans$ & $3.39(0.201)$ & $3.87(0.016)$ & $4.26(0.005)$ & $4.56(0.059)$ & $1801$\\
6 $cis$   & $3.50(0.219)$ & $3.86(0.017)$ & $4.41(0.006)$ & $4.51(0.000)$ & $1809$\\
7 $trans$ & $3.19(0.219)$ & $3.74(0.001)$ & $4.37(0.006)$ & $4.38(0.002)$ & $1941$\\
7 $cis$   & $3.36(0.205)$ & $3.76(0.004)$ & $4.42(0.001)$ & $4.48(0.017)$ & $1982$\\
8 $trans$ & $3.73(0.134)$ & $4.21(0.001)$ & $4.43(0.002)$ & $4.58(0.033)$ & $1911$\\
8 $cis$   & $3.72(0.169)$ & $4.22(0.000)$ & $4.30(0.004)$ & $4.47(0.026)$ & $1916$\\
9 $trans$ & $3.49(0.182)$ & $3.96(0.003)$ & $4.49(0.004)$ & $4.58(0.004)$ & $1902$\\
9 $cis$   & $3.47(0.202)$ & $3.89(0.013)$ & $4.47(0.091)$ & $4.49(0.002)$ & $1838$\\
10 $trans$ & $3.19(0.219)$ & $4.20(0.000)$ & $4.31(0.000)$ & $4.34(0.003)$ & $1761$\\
10 $cis$   & $3.31(0.229)$ & $4.20(0.000)$ & $4.37(0.002)$ & $4.66(0.411)$ & $1787$\\
11 $trans$ & $3.41(0.175)$ & $3.91(0.000)$ & $4.44(0.006)$ & $4.59(0.000)$ & $1679$\\
11 $cis$   & $3.13(0.263)$ & $3.92(0.000)$ & $4.38(0.000)$ & $4.67(0.416)$ & $1839$\\
12 $trans$ & $3.41(0.218)$ & $3.85(0.000)$ & $4.47(0.003)$ & $4.54(0.000)$ & $1725$\\
12 $cis$   & $3.54(0.225)$ & $3.86(0.000)$ & $4.54(0.000)$ & $4.55(0.001)$ & $1725$\\
\hline
\end{tabular}
\end{table}

\newpage

\begin{table}
\caption{Vertical excitation energies of four lowest excited states, corresponding oscillator strengths from the ground electronic state, and Grimme-D3(BJ) ground state Dispersion energies in $\textit{trans}$ and $\textit{cis}$ NADH conformations in methanol. }
\label{tab:nadh_ex_energy_methanol}
\begin{tabular}{c c c c c c}
\hline
  & Energy, eV    & Energy, eV.   & Energy, eV    & Energy, eV    & Dispersion\\
Conformation    & $(Osc. str.)$ & $(Osc. str.)$ & $(Osc. str.)$ & $(Osc. str.)$ & energy,\\
  & Ex. state 1       & Ex. state 2       & Ex. state 3       & Ex. state 4       & Hartree, 10$^{-4}$\\
\hline
1 $trans$ & $3.13(0.156)$ & $3.67(0.019)$ & $3.95(0.029)$ & $4.23(0.009)$ & $2023$\\
1 $cis$   & $3.27(0.202)$ & $3.76(0.001)$ & $3.96(0.012)$ & $4.35(0.008)$ & $2003$\\
2 $trans$ & $3.56(0.152)$ & $3.74(0.012)$ & $4.36(0.006)$ & $4.41(0.007)$ & $1887$\\
2 $cis$   & $3.39(0.119)$ & $3.71(0.045)$ & $4.29(0.022)$ & $4.31(0.021)$ & $1921$\\
3 $trans$ & $3.45(0.147)$ & $3.49(0.023)$ & $4.23(0.001)$ & $4.52(0.038)$ & $1965$\\
3 $cis$   & $3.22(0.037)$ & $3.52(0.149)$ & $3.99(0.002)$ & $4.39(0.181)$ & $1971$\\
4 $trans$ & $3.14(0.146)$ & $3.42(0.034)$ & $4.11(0.004)$ & $4.33(0.064)$ & $1965$\\
4 $cis$   & $3.18(0.159)$ & $3.50(0.039)$ & $4.16(0.002)$ & $4.26(0.054)$ & $1956$\\
5 $trans$ & $3.44(0.148)$ & $3.73(0.009)$ & $4.32(0.007)$ & $4.54(0.127)$ & $2017$\\
5 $cis$   & $3.57(0.106)$ & $3.65(0.070)$ & $4.28(0.002)$ & $4.52(0.001)$ & $1971$\\
6 $trans$ & $3.40(0.197)$ & $3.88(0.017)$ & $4.25(0.005)$ & $4.57(0.110)$ & $1802$\\
6 $cis$   & $3.51(0.215)$ & $3.87(0.017)$ & $4.39(0.006)$ & $4.51(0.000)$ & $1811$\\
7 $trans$ & $3.19(0.215)$ & $3.75(0.001)$ & $4.36(0.006)$ & $4.38(0.001)$ & $1942$\\
7 $cis$   & $3.36(0.201)$ & $3.76(0.004)$ & $4.42(0.001)$ & $4.47(0.016)$ & $1983$\\
8 $trans$ & $3.74(0.131)$ & $4.22(0.001)$ & $4.43(0.002)$ & $4.58(0.032)$ & $1912$\\
8 $cis$   & $3.74(0.166)$ & $4.23(0.000)$ & $4.29(0.004)$ & $4.46(0.023)$ & $1917$\\
9 $trans$ & $3.49(0.179)$ & $3.98(0.003)$ & $4.48(0.004)$ & $4.57(0.000)$ & $1904$\\
9 $cis$   & $3.47(0.193)$ & $3.92(0.015)$ & $4.43(0.101)$ & $4.52(0.001)$ & $1855$\\
10 $trans$ & $3.19(0.215)$ & $4.22(0.000)$ & $4.28(0.000)$ & $4.33(0.003)$ & $1762$\\
10 $cis$   & $3.31(0.225)$ & $4.23(0.000)$ & $4.34(0.002)$ & $4.67(0.405)$ & $1787$\\
11 $trans$ & $3.41(0.172)$ & $3.91(0.000)$ & $4.43(0.005)$ & $4.59(0.000)$ & $1678$\\
11 $cis$   & $3.13(0.259)$ & $3.89(0.000)$ & $4.41(0.000)$ & $4.67(0.409)$ & $1840$\\
12 $trans$ & $3.42(0.216)$ & $3.84(0.000)$ & $4.47(0.003)$ & $4.53(0.000)$ & $1725$\\
12 $cis$   & $3.54(0.224)$ & $3.85(0.000)$ & $4.54(0.000)$ & $4.55(0.001)$ & $1725$\\
\hline
\end{tabular}
\end{table}

Transition dipole moment (TDM) components under excitation of several NADH conformers from their ground state to the first excited state in aqueous and methanol solutions are presented in Tables \ref{tab:TDM_xyz_angle_water} and \ref{tab:TDM_xyz_angle_methanol}, respectively.

Axis Z in Tables \ref{tab:TDM_xyz_angle_water} and \ref{tab:TDM_xyz_angle_methanol} is parallel to the direction ($C6N \rightarrow C3N$) and axis Y is directed along the vector product ($C6N \rightarrow C3N) \times (C4N \rightarrow C2N$) that is perpendicular to the NA plane in the case of negligible puckering effect. All directions are shown in Fig.\ref{fig:NADHdist}.

\newpage
\begin{table}
\caption{TDM components from the ground electronic state to the first excited state in 24 NADH conformations in aqueous solution. $\gamma$ is an angle between TDM direction and the NA ring plane. }
\label{tab:TDM_xyz_angle_water}
\begin{tabular}{c c c c c}
\hline
Conformation & Z  & X & Y    & $ \gamma$, degrees\\
\hline
1 $trans$ & $-1.384$ & $0.010$ & $0.161$ & $6.65$\\
1 $cis$ & $1.544$ & $0.036$ & $-0.125$ & $-4.64$\\
2 $trans$ & $-1.260$ & $0.171$ & $0.160$ & $7.16$\\
2 $cis$ & $-1.154$ & $0.109$ & $0.129$ & $6.33$\\
3 $trans$ & $-1.299$ & $-0.060$ & $0.043$ & $1.89$\\
3 $cis$ & $-0.597$ & $0.114$ & $0.318$ & $27.57$\\
4 $trans$ & $-1.337$ & $0.028$ & $0.174$ & $7.39$\\
4 $cis$ & $-1.398$ & $0.064$ & $0.074$ & $3.02$\\
5 $trans$ & $-1.149$ & $0.014$ & $0.066$ & $3.30$\\
5 $cis$ & $-1.134$ & $0.034$ & $-0.064$ & $-3.25$\\
6 $trans$ & $-1.498$ & $0.053$ & $-0.084$ & $-3.22$\\
6 $cis$ & $-1.540$ & $0.046$ & $-0.108$ & $-4.02$\\
7 $trans$ & $-1.611$ & $0.001$ & $0.081$ & $2.86$\\
7 $cis$ & $-1.519$ & $0.053$ & $0.086$ & $3.24$\\
8 $trans$ & $-1.172$ & $0.014$ & $0.038$ & $1.85$\\
8 $cis$ & $-1.312$ & $-0.033$ & $0.091$ & $3.95$\\
9 $trans$ & $-1.408$ & $0.044$ & $0.015$ & $0.61$\\
9 $cis$ & $-1.484$ & $0.003$ & $-0.091$ & $-3.51$\\
10 $trans$ & $-1.612$ & $0.033$ & $0.074$ & $2.63$\\
10 $cis$ & $-1.622$ & $-0.038$ & $0.027$ & $0.95$\\
11 $trans$ & $-1.390$ & $0.063$ & $0.111$ & $4.55$\\
11 $cis$ & $-1.760$ & $0.016$ & $0.288$ & $9.30$\\
12 $trans$ & $-1.557$ & $0.016$ & $-0.025$ & $-0.92$\\
12 $cis$ & $-1.557$ & $-0.020$ & $-0.027$ & $-0.98$\\
\hline
\end{tabular}
\end{table}

\begin{table}
\caption{TDM components from the ground electronic state to the first excited state in 24 NADH conformations in methanol solution. $\gamma$ is an angle between TDM direction and the NA ring plane. }
\label{tab:TDM_xyz_angle_methanol}
\begin{tabular}{c c c c c}
\hline
Conformation & Z  & X & Y    & $ \gamma$, degrees\\
\hline
1 $trans$ & $-1.361$ & $0.006$ & $0.162$ & $6.77$\\
1 $cis$ & $1.523$ & $0.041$ & $-0.123$ & $-4.61$\\
2 $trans$ & $-1.254$ & $0.174$ & $0.155$ & $6.97$\\
2 $cis$ & $-1.145$ & $0.115$ & $0.127$ & $6.29$\\
3 $trans$ & $-1.270$ & $-0.060$ & $0.038$ & $1.71$\\
3 $cis$ & $-0.571$ & $0.114$ & $0.317$ & $28.59$\\
4 $trans$ & $-1.315$ & $0.027$ & $0.172$ & $7.45$\\
4 $cis$ & $-1.375$ & $0.062$ & $0.072$ & $2.99$\\
5 $trans$ & $-1.279$ & $0.034$ & $0.076$ & $3.42$\\
5 $cis$ & $-1.061$ & $0.026$ & $-0.064$ & $-3.44$\\
6 $trans$ & $-1.482$ & $0.057$ & $-0.086$ & $-3.32$\\
6 $cis$ & $-1.522$ & $0.049$ & $-0.107$ & $-4.02$\\
7 $trans$ & $-1.592$ & $0.001$ & $0.082$ & $2.95$\\
7 $cis$ & $-1.503$ & $0.052$ & $0.086$ & $3.26$\\
8 $trans$ & $-1.156$ & $0.010$ & $0.037$ & $1.84$\\
8 $cis$ & $-1.299$ & $-0.038$ & $0.088$ & $3.88$\\
9 $trans$ & $-1.396$ & $0.046$ & $0.013$ & $0.55$\\
9 $cis$ & $-1.454$ & $0.013$ & $-0.054$ & $-2.15$\\
10 $trans$ & $-1.596$ & $0.031$ & $0.072$ & $2.60$\\
10 $cis$ & $-1.604$ & $-0.044$ & $0.027$ & $0.98$\\
11 $trans$ & $-1.380$ & $0.062$ & $0.113$ & $4.68$\\
11 $cis$ & $-1.742$ & $0.014$ & $0.289$ & $9.42$\\
12 $trans$ & $-1.549$ & $0.017$ & $-0.022$ & $-0.82$\\
12 $cis$ & $-1.550$ & $-0.019$ & $-0.026$ & $-0.97$\\
\hline
\end{tabular}
\end{table}

\newpage

\begin{table}
\caption{TDM fluorescence components from the first excited state to the ground electronic state in the NADH conformations {relaxed}  in aqueous solution. $\gamma$ is an angle between the fluorescence TDM direction and the NA ring plane.  $\xi$ is an angle between the excitation and fluorescence TDM directions.}
\label{tab:TDM_xyz_angle_water1}
\begin{tabular}{c c c c c c}
\hline
Conformation & Z$_r$  & X$_r$ & Y$_r$    & $\gamma$, degrees & $\xi$, dergees\\
\hline
10 $trans$ \:  & 1.136 & 0.110 & -0.076 & -3.79 & 6.79 \\
11 $trans$ \:  & 1.105 & 0.007 & -0.071 & -3.67 & 6.07 \\
11 $cis$ \:  & 1.764 & 0.025 & -0.230 & 7.44 & 2.28 \\
\hline
\end{tabular}
\end{table}

Note that the coordinates X, Y, and Z in Tables \ref{tab:TDM_xyz_angle_water} and \ref{tab:TDM_xyz_angle_methanol} refer to the NADH ground state nuclear configuration, while the coordinates  X$_r$, Y$_r$, and Z$_r$ in Table \ref{tab:TDM_xyz_angle_water1} refer to the NADH excited state relaxed nuclear configuration.

Bonging and dihedral angles describing the deformations in the ground and first excited states of the NA moiety in the conformation 11 of NADH (see Table \ref{tab:nadh_conf_geom_water})  in water are collected in Tables \ref{table:11BN} and \ref{table:11DH}. The designation of atoms follows Fig.\ref{fig:ring}.

\begin{table}
  \caption{Bonding angles describing the deformations of the NA moiety in the conformation 11 of NADH in water in the ground and first excited states. }
  \label{table:11BN}
\begin{tabular}{ccccc}
\hline
Bonding angle & Ground state, & Ground state, & Excited state, & Excited state, \\
   & \emph{cis}, degrees & \emph{trans}, degrees & \emph{cis}, degrees & \emph{trans}, degrees \\ \hline
C3-C7-O & 126.5 & 120.6 & 125.2 & 122.9 \\ \hline
C4-C3-C7 & 120.8 & 116.3 & 118.1 & 116.0 \\ \hline
C2-C3-C7 & 118.0 & 121.7 & 121.6 & 124.8  \\
\hline
\end{tabular}
\end{table}

\begin{table}
  \caption{Dihedral angles describing the deformations of the NA moiety in the conformation 11 of NADH in water in the ground and first excited states. }
  \label{table:11DH}
\begin{tabular}{ccccc}
\hline
Dihedral angle & Ground state, & Ground state, & Excited state, & Excited state, \\
   & \emph{cis}, degrees & \emph{trans}, degrees  & \emph{cis}, degrees & \emph{trans}, degrees \\ \hline
N-C2-C3-C4 & -2.0 & -4.4 & -5.0 & 9.3  \\ \hline
C2-C3-C4-C5 & 3.5 &  7.1 &  6.1 &-2.8  \\ \hline
C3-C4-C5-C6 & -3.0& -5.9 & -3.5 &-3.9  \\ \hline
C4-C5-C6-N  & 0.9 &  1.9 & -0.5 & 4.2  \\ \hline
C5-C6-N-C2  & 1.0 &  1.8 &  2.1 & 2.5  \\ \hline
C6-N-C2-C3  & -0.5& -0.5 &  0.7 &-9.4  \\
\hline
\end{tabular}
\end{table}

\section*{Derivation of eq.(\ref{eq:expansion})}
We consider the anisotropic part of the experimental signal in eq.(\ref{eq:aniso}) consisting of a sum of two unresolved exponents:
\begin{eqnarray}
\label{eq:two_exp}
I_{aniso}(t)=N_{fol} w_{fol} exp(-k_{fol} t) + (1-N_{fol}) w_{un} exp(-k_{un} t),
\end{eqnarray}
where $k_{fol}=\tau^{-1}_{fol}$ and $k_{un}=\tau^{-1}_{un}$ are rotational diffusion rate constants for folded and unfolded confirmations, respectively, $N_{fol}$ is the relative ground state concentration  of the folded conformation that is a function of the solution viscosity $\eta$, and the terms $w_{fol}$ and $w_{un}$ are coefficients accounting for excitation probabilities from the ground state to the first excited state and transformation of the molecular nuclear configuration due to fast vibrational relaxation after excitation by the laser pulse.

The relationship between the probabilities $w_{fol}$ and $w_{un}$ can be obtained using the results reported by Freed et al. \cite{Freed1967} who experimentally determined the fluorescence intensity in NADH under excitation at 334~nm in methanol-water solution as a function of methanol concentration $I_{fl}(\eta)$. This relationship can be presented in a form:
\begin{eqnarray}
\label{eq:two_exp_steady}
I_{fl}(\eta)=N_{fol} w_{fol} + (1-N_{fol}) w_{un},
\end{eqnarray}
where the probability $w_{un}$ can be determined from the fluorescence intensity in pure methanol $w_{un}=I_{fl}(M)$.

In general, the rate constants $k_{fol}$ and $k_{un}$ are not the same and the corresponding signal has a double exponential form. However, if $(k_{fol}-k_{un}) t \ll 1$, the experimental signal in eq.(\ref{eq:two_exp}) can be approximately presented in a single exponential form:
\begin{eqnarray}
\label{eq:single_exp}
I_{aniso}(t)=N_{fol} w_{fol}\, exp(-k_{fol} t) + (1-N_{fol}) w_{un}\,exp(-k_{un} t) \approx C\, exp(-k t),
\end{eqnarray}
where the constant $C$ is equal to $C=N_{fol} w_{fol}+(1-N_{fol}) w_{un}$ as eq.(\ref{eq:single_exp}) must be fulfilled at $t=0$.

Using the Taylor expansion the sum of two exponents in eq.(\ref{eq:single_exp}) can be transformed as:
\begin{eqnarray}
\label{eq:double}
&&C exp(-\gamma t) \approx
 C e^{-k_{fol} t}\,\left( 1 -  \frac{(1-N_{fol})w_{un}}{C}\,((k_{un}-k_{fol}) t+\frac{1}{2}(k_{un}-k_{fol})^2 t^2) \right).
\end{eqnarray}

Taking the logarithm from left and right parts in eq.(\ref{eq:double}) and keeping only the linear on $(k_{un}-k_{fol})$ terms one gets:
\begin{eqnarray}
\label{eq:k}
&&k  \approx
  \frac{N_{fol} w_{fol}}{C}k_{fol}   + \frac{(1-N_{fol})w_{un}}{C}\,k_{un}.
\end{eqnarray}

Equation (\ref{eq:k}) can be readily transformed to the form of eq.(\ref{eq:expansion}).

\end{suppinfo}

\newpage

\bibliography{NADH_Fluo2020}

\providecommand{\latin}[1]{#1}
\makeatletter
\providecommand{\doi}
  {\begingroup\let\do\@makeother\dospecials
  \catcode`\{=1 \catcode`\}=2 \doi@aux}
\providecommand{\doi@aux}[1]{\endgroup\texttt{#1}}
\makeatother
\providecommand*\mcitethebibliography{\thebibliography}
\csname @ifundefined\endcsname{endmcitethebibliography}
  {\let\endmcitethebibliography\endthebibliography}{}
\begin{mcitethebibliography}{61}
\providecommand*\natexlab[1]{#1}
\providecommand*\mciteSetBstSublistMode[1]{}
\providecommand*\mciteSetBstMaxWidthForm[2]{}
\providecommand*\mciteBstWouldAddEndPuncttrue
  {\def\EndOfBibitem{\unskip.}}
\providecommand*\mciteBstWouldAddEndPunctfalse
  {\let\EndOfBibitem\relax}
\providecommand*\mciteSetBstMidEndSepPunct[3]{}
\providecommand*\mciteSetBstSublistLabelBeginEnd[3]{}
\providecommand*\EndOfBibitem{}
\mciteSetBstSublistMode{f}
\mciteSetBstMaxWidthForm{subitem}{(\alph{mcitesubitemcount})}
\mciteSetBstSublistLabelBeginEnd
  {\mcitemaxwidthsubitemform\space}
  {\relax}
  {\relax}

\bibitem[Chance \latin{et~al.}(1962)Chance, Cohen, Jobsis, and
  Schoen]{Chance1962}
Chance,~B.; Cohen,~P.; Jobsis,~F.; Schoen,~B. {Intracellular Oxidatior
  Reduction States in Vivo}. \emph{Science} \textbf{1962}, \emph{137},
  499--508\relax
\mciteBstWouldAddEndPuncttrue
\mciteSetBstMidEndSepPunct{\mcitedefaultmidpunct}
{\mcitedefaultendpunct}{\mcitedefaultseppunct}\relax
\EndOfBibitem
\bibitem[Van~der Heiden \latin{et~al.}(2009)Van~der Heiden, Cantley, and
  Thompson]{Heiden2009}
Van~der Heiden,~M.~G.; Cantley,~L.; Thompson,~C.~B. \emph{Science}
  \textbf{2009}, \emph{324}, 1029\relax
\mciteBstWouldAddEndPuncttrue
\mciteSetBstMidEndSepPunct{\mcitedefaultmidpunct}
{\mcitedefaultendpunct}{\mcitedefaultseppunct}\relax
\EndOfBibitem
\bibitem[Belenky \latin{et~al.}(2007)Belenky, Bogan, and Brenner]{Belenky2007}
Belenky,~P.; Bogan,~K.~L.; Brenner,~C. {NAD+ Metabolism in Health and Disease}.
  \emph{Trends Biochem. Sci.} \textbf{2007}, \emph{32}, 12--19\relax
\mciteBstWouldAddEndPuncttrue
\mciteSetBstMidEndSepPunct{\mcitedefaultmidpunct}
{\mcitedefaultendpunct}{\mcitedefaultseppunct}\relax
\EndOfBibitem
\bibitem[Pollak \latin{et~al.}(2007)Pollak, D{\"{o}}lle, and
  Ziegler]{Pollak2007}
Pollak,~N.; D{\"{o}}lle,~C.; Ziegler,~M. {The power to reduce: Pyridine
  nucleotides - Small molecules with a multitude of functions}. \emph{Biochem.
  J.} \textbf{2007}, \emph{402}, 205--218\relax
\mciteBstWouldAddEndPuncttrue
\mciteSetBstMidEndSepPunct{\mcitedefaultmidpunct}
{\mcitedefaultendpunct}{\mcitedefaultseppunct}\relax
\EndOfBibitem
\bibitem[{Da Veiga Moreira} \latin{et~al.}(2016){Da Veiga Moreira}, Hamraz,
  Abolhassani, Bigan, P{\'{e}}r{\`{e}}s, Paulev{\'{e}}, Nogueira, Steyaert, and
  Schwartz]{Moreira2016}
{Da Veiga Moreira},~J.; Hamraz,~M.; Abolhassani,~M.; Bigan,~E.;
  P{\'{e}}r{\`{e}}s,~S.; Paulev{\'{e}},~L.; Nogueira,~M.~L.; Steyaert,~J.~M.;
  Schwartz,~L. {The redox status of cancer cells supports mechanisms behind the
  Warburg effect}. \emph{Metabolites} \textbf{2016}, \emph{6}, 1--12\relax
\mciteBstWouldAddEndPuncttrue
\mciteSetBstMidEndSepPunct{\mcitedefaultmidpunct}
{\mcitedefaultendpunct}{\mcitedefaultseppunct}\relax
\EndOfBibitem
\bibitem[Barlow and Chance(1976)Barlow, and Chance]{Barlow1976}
Barlow,~C.~H.; Chance,~B. {Ischemic areas in perfused rat hearts: Measurement
  by NADH fluorescence photography}. \emph{Science} \textbf{1976}, \emph{193},
  909--910\relax
\mciteBstWouldAddEndPuncttrue
\mciteSetBstMidEndSepPunct{\mcitedefaultmidpunct}
{\mcitedefaultendpunct}{\mcitedefaultseppunct}\relax
\EndOfBibitem
\bibitem[Kasischke \latin{et~al.}(2004)Kasischke, Vishwasrao, Fisher, Zipfel,
  and Webb]{Kasischke2004}
Kasischke,~K.~A.; Vishwasrao,~H.~D.; Fisher,~P.~J.; Zipfel,~W.~R.; Webb,~W.~W.
  {Neural activity triggers neuronal oxidative metabolism followed by
  astrocytic glycolysis}. \emph{Science} \textbf{2004}, \emph{305},
  99--103\relax
\mciteBstWouldAddEndPuncttrue
\mciteSetBstMidEndSepPunct{\mcitedefaultmidpunct}
{\mcitedefaultendpunct}{\mcitedefaultseppunct}\relax
\EndOfBibitem
\bibitem[Blinova \latin{et~al.}(2005)Blinova, Carroll, Bose, Smirnov, Harvey,
  Knutson, and Balaban]{Blinova2005}
Blinova,~K.; Carroll,~S.; Bose,~S.; Smirnov,~A.~V.; Harvey,~J.~J.;
  Knutson,~J.~R.; Balaban,~R.~S. {Distribution of mitochondrial NADH
  fluorescence lifetimes: Steady-state kinetics of matrix NADH interactions}.
  \emph{Biochemistry} \textbf{2005}, \emph{44}, 2585--2594\relax
\mciteBstWouldAddEndPuncttrue
\mciteSetBstMidEndSepPunct{\mcitedefaultmidpunct}
{\mcitedefaultendpunct}{\mcitedefaultseppunct}\relax
\EndOfBibitem
\bibitem[Lakowicz \latin{et~al.}(1992)Lakowicz, Szmacinski, Nowaczyk, and
  Johnson]{Lakowicz1992}
Lakowicz,~J.~R.; Szmacinski,~H.; Nowaczyk,~K.; Johnson,~M.~L. {F}luorescence
  lifetime imaging of free and protein-bound {NADH}. \emph{Proc. Natl. Acad.
  Sci. USA} \textbf{1992}, \emph{89}, 1271--1275\relax
\mciteBstWouldAddEndPuncttrue
\mciteSetBstMidEndSepPunct{\mcitedefaultmidpunct}
{\mcitedefaultendpunct}{\mcitedefaultseppunct}\relax
\EndOfBibitem
\bibitem[Yaseen \latin{et~al.}(2017)Yaseen, Sutin, Wu, Fu, Uhlirova, Devor,
  Boas, and Sakad{\v{z}}i{\'{c}}]{Yaseen2017}
Yaseen,~M.~A.; Sutin,~J.; Wu,~W.; Fu,~B.; Uhlirova,~H.; Devor,~A.; Boas,~D.~A.;
  Sakad{\v{z}}i{\'{c}},~S. {Fluorescence lifetime microscopy of NADH
  distinguishes alterations in cerebral metabolism in vivo}. \emph{Biom. Opt.
  Express} \textbf{2017}, \emph{8}, 2368--2385\relax
\mciteBstWouldAddEndPuncttrue
\mciteSetBstMidEndSepPunct{\mcitedefaultmidpunct}
{\mcitedefaultendpunct}{\mcitedefaultseppunct}\relax
\EndOfBibitem
\bibitem[Evers \latin{et~al.}(2018)Evers, Salma, Osseiran, Casper, Birngruber,
  Evans, and Manstein]{Evers2018}
Evers,~M.; Salma,~N.; Osseiran,~S.; Casper,~M.; Birngruber,~R.; Evans,~C.~L.;
  Manstein,~D. {Enhanced quantification of metabolic activity for individual
  adipocytes by label-free FLIM}. \emph{Sci. Reports} \textbf{2018}, \emph{8},
  1--14\relax
\mciteBstWouldAddEndPuncttrue
\mciteSetBstMidEndSepPunct{\mcitedefaultmidpunct}
{\mcitedefaultendpunct}{\mcitedefaultseppunct}\relax
\EndOfBibitem
\bibitem[Schaefer \latin{et~al.}(2019)Schaefer, Kalinina, Rueck, von Arnim, and
  von Einem]{Schaefer2019}
Schaefer,~P.~M.; Kalinina,~S.; Rueck,~A.; von Arnim,~C.~A.; von Einem,~B. {NADH
  Autofluorescence—A Marker on its Way to Boost Bioenergetic Research}.
  \emph{Cytometry Part A} \textbf{2019}, \emph{95}, 34--46\relax
\mciteBstWouldAddEndPuncttrue
\mciteSetBstMidEndSepPunct{\mcitedefaultmidpunct}
{\mcitedefaultendpunct}{\mcitedefaultseppunct}\relax
\EndOfBibitem
\bibitem[Konig \latin{et~al.}(1997)Konig, Berns, and Tromberg]{Konig1997}
Konig,~K.; Berns,~M.~W.; Tromberg,~B.~J. {Time-resolved and steady-state
  fluorescence measurements of $\beta$-nicotinamide adenine
  dinucleotide-alcohol dehydrogenase complex during UVA exposure}. \emph{J.
  Photochem. Photobiol. B: Biology} \textbf{1997}, \emph{37}, 91--95\relax
\mciteBstWouldAddEndPuncttrue
\mciteSetBstMidEndSepPunct{\mcitedefaultmidpunct}
{\mcitedefaultendpunct}{\mcitedefaultseppunct}\relax
\EndOfBibitem
\bibitem[Vishwasrao \latin{et~al.}(2005)Vishwasrao, Heikal, Kasischke, and
  Webb]{Vishwasrao2005}
Vishwasrao,~H.~D.; Heikal,~A.~A.; Kasischke,~K.~A.; Webb,~W.~W. {Conformational
  dependence of intracellular NADH on metabolic state revealed by associated
  fluorescence anisotropy}. \emph{J. Biol. Chem.} \textbf{2005}, \emph{280},
  25119--25126\relax
\mciteBstWouldAddEndPuncttrue
\mciteSetBstMidEndSepPunct{\mcitedefaultmidpunct}
{\mcitedefaultendpunct}{\mcitedefaultseppunct}\relax
\EndOfBibitem
\bibitem[Visser and van Hoek(1981)Visser, and van Hoek]{Visser1981}
Visser,~A. J. W.~G.; van Hoek,~A. {T}he fluorescence decay of reduced
  nicotinamides in aqueous solution after excitation with a UV--mode locked
  {A}r ion laser. \emph{Photochem. Photobiol.} \textbf{1981}, \emph{33},
  35--40\relax
\mciteBstWouldAddEndPuncttrue
\mciteSetBstMidEndSepPunct{\mcitedefaultmidpunct}
{\mcitedefaultendpunct}{\mcitedefaultseppunct}\relax
\EndOfBibitem
\bibitem[Couprie \latin{et~al.}(1994)Couprie, Merola, Tauc, Garzella,
  Delboulb{\'{e}}, Hara, and Billardon]{Couprie1994a}
Couprie,~M.~E.; Merola,~F.; Tauc,~P.; Garzella,~D.; Delboulb{\'{e}},~A.;
  Hara,~T.; Billardon,~M. {First use of the UV Super-ACO free-electron laser:
  Fluorescence decays and rotational dynamics of the NADH coenzyme}. \emph{Rev.
  Sci. Instr.} \textbf{1994}, \emph{65}, 1485--1495\relax
\mciteBstWouldAddEndPuncttrue
\mciteSetBstMidEndSepPunct{\mcitedefaultmidpunct}
{\mcitedefaultendpunct}{\mcitedefaultseppunct}\relax
\EndOfBibitem
\bibitem[Hull \latin{et~al.}(2001)Hull, Conger, and Hoobler]{Hull2001}
Hull,~R.~V.; Conger,~P.~S.; Hoobler,~R.~J. {Conformation of NADH studied by
  fluorescence excitation transfer spectroscopy}. \emph{Biophys. Chem.}
  \textbf{2001}, \emph{90}, 9--16\relax
\mciteBstWouldAddEndPuncttrue
\mciteSetBstMidEndSepPunct{\mcitedefaultmidpunct}
{\mcitedefaultendpunct}{\mcitedefaultseppunct}\relax
\EndOfBibitem
\bibitem[Ladokhin and Brand(1995)Ladokhin, and Brand]{Ladokhin1995}
Ladokhin,~A.~S.; Brand,~L. {Evidence for an excited-state reaction contributing
  to NADH fluorescence}. \emph{J. Fluores.} \textbf{1995}, \emph{5},
  99--106\relax
\mciteBstWouldAddEndPuncttrue
\mciteSetBstMidEndSepPunct{\mcitedefaultmidpunct}
{\mcitedefaultendpunct}{\mcitedefaultseppunct}\relax
\EndOfBibitem
\bibitem[Blacker \latin{et~al.}(2013)Blacker, Marsh, Duchen, and
  Bain]{Blacker2013}
Blacker,~T.~S.; Marsh,~R.~J.; Duchen,~M.~R.; Bain,~A.~J. {Activated barrier
  crossing dynamics in the non-radiative decay of NADH and NADPH}. \emph{Chem.
  Phys.} \textbf{2013}, \emph{422}, 184--194\relax
\mciteBstWouldAddEndPuncttrue
\mciteSetBstMidEndSepPunct{\mcitedefaultmidpunct}
{\mcitedefaultendpunct}{\mcitedefaultseppunct}\relax
\EndOfBibitem
\bibitem[Oppenheimer \latin{et~al.}(1971)Oppenheimer, Arnold, and
  Kaplan]{Oppenheimer1971}
Oppenheimer,~N.~J.; Arnold,~L.~J.; Kaplan,~N.~O. {A structure of pyridine
  nucleotides in solution.} \emph{Proc. Nat. Acad. Sci. USA} \textbf{1971},
  \emph{68}, 3200--3205\relax
\mciteBstWouldAddEndPuncttrue
\mciteSetBstMidEndSepPunct{\mcitedefaultmidpunct}
{\mcitedefaultendpunct}{\mcitedefaultseppunct}\relax
\EndOfBibitem
\bibitem[McDonald \latin{et~al.}(1972)McDonald, Brown, Hollis, and
  Walter]{McDonald1972}
McDonald,~G.; Brown,~B.; Hollis,~D.; Walter,~C. {Some Effects of Environment on
  the Folding of Nicotinamide-Adenine Dinucleotides in Aqueous Solutions}.
  \emph{Biochemistry} \textbf{1972}, \emph{11}, 1920--1930\relax
\mciteBstWouldAddEndPuncttrue
\mciteSetBstMidEndSepPunct{\mcitedefaultmidpunct}
{\mcitedefaultendpunct}{\mcitedefaultseppunct}\relax
\EndOfBibitem
\bibitem[Babu and Lim(2016)Babu, and Lim]{Babu2016}
Babu,~C.~S.; Lim,~C. {Efficient Binding of Flexible and Redox-Active Coenzymes
  by Oxidoreductases}. \emph{ACS Catalysis} \textbf{2016}, \emph{6},
  3469--3472\relax
\mciteBstWouldAddEndPuncttrue
\mciteSetBstMidEndSepPunct{\mcitedefaultmidpunct}
{\mcitedefaultendpunct}{\mcitedefaultseppunct}\relax
\EndOfBibitem
\bibitem[Cao \latin{et~al.}(2019)Cao, Wu, Zhang, and Dolg]{Dolg2019}
Cao,~X.; Wu,~L.; Zhang,~J.; Dolg,~M. {Density Functional Studies of Coenzyme
  NADPH and Its Oxidized Form NADP$^+$: Structures, UV–Vis Spectra, and the
  Oxidation Mechanism of NADPH}. \emph{J. Comput. Chem.} \textbf{2019},
  \emph{9999}, 1--12\relax
\mciteBstWouldAddEndPuncttrue
\mciteSetBstMidEndSepPunct{\mcitedefaultmidpunct}
{\mcitedefaultendpunct}{\mcitedefaultseppunct}\relax
\EndOfBibitem
\bibitem[Cadena-Caicedo \latin{et~al.}(2020)Cadena-Caicedo, Gonzalez-Cano,
  L{\'{o}}pez-Arteaga, Esturau-Escofet, and Peon]{Peon2020}
Cadena-Caicedo,~A.; Gonzalez-Cano,~B.; L{\'{o}}pez-Arteaga,~R.;
  Esturau-Escofet,~N.; Peon,~J. {Ultrafast Fluorescence Signals from NADH:
  Resonant Energy Transfer in the Folded and Unfolded Forms}. \emph{J. Phys.
  Chem. B} \textbf{2020}, \emph{124}, 519--530\relax
\mciteBstWouldAddEndPuncttrue
\mciteSetBstMidEndSepPunct{\mcitedefaultmidpunct}
{\mcitedefaultendpunct}{\mcitedefaultseppunct}\relax
\EndOfBibitem
\bibitem[Freed \latin{et~al.}(1967)Freed, Neyfakh, and Tumerman]{Freed1967}
Freed,~S.; Neyfakh,~E.; Tumerman,~L. {Influence of solvents on the
  intramolecular energy transfer in NADH and NADPH}. \emph{Biochim. Biophys.
  Acta} \textbf{1967}, \emph{143}, 432--434\relax
\mciteBstWouldAddEndPuncttrue
\mciteSetBstMidEndSepPunct{\mcitedefaultmidpunct}
{\mcitedefaultendpunct}{\mcitedefaultseppunct}\relax
\EndOfBibitem
\bibitem[Heiner \latin{et~al.}(2017)Heiner, Roland, Leonard, Haacke, and
  Groma]{Heiner2017}
Heiner,~Z.; Roland,~T.; Leonard,~J.; Haacke,~S.; Groma,~G.~I. {Kinetics of
  Light-Induced Intramolecular Energy Transfer in Different Conformational
  States of NADH}. \emph{J. Phys. Chem. B} \textbf{2017}, \emph{121},
  8037--8045\relax
\mciteBstWouldAddEndPuncttrue
\mciteSetBstMidEndSepPunct{\mcitedefaultmidpunct}
{\mcitedefaultendpunct}{\mcitedefaultseppunct}\relax
\EndOfBibitem
\bibitem[Sasin \latin{et~al.}(2020)Sasin, Gorbunova, and
  Vasyutinskii]{Gorbunova20}
Sasin,~M.~E.; Gorbunova,~I.; Vasyutinskii,~O.~S. Observation of Anisotropic
  Relaxation in Biological Molecules with Subpicosecond Temporal Resolution.
  \emph{Tech. Phys. Lett.} \textbf{2020}, \emph{46}, 158--160\relax
\mciteBstWouldAddEndPuncttrue
\mciteSetBstMidEndSepPunct{\mcitedefaultmidpunct}
{\mcitedefaultendpunct}{\mcitedefaultseppunct}\relax
\EndOfBibitem
\bibitem[Gorbunova \latin{et~al.}(2020)Gorbunova, Sasin, Beltukov, Semenov, and
  Vasyutinskii]{Gorbunova20b}
Gorbunova,~I.~A.; Sasin,~M.~E.; Beltukov,~Y.~M.; Semenov,~A.~A.;
  Vasyutinskii,~O.~S. Anisotropic relaxation in NADH excited states studied by
  polarization-modulation pump-probe transient spectroscopy. \emph{Phys. Chem.
  Chem. Phys.} \textbf{2020}, \relax
\mciteBstWouldAddEndPunctfalse
\mciteSetBstMidEndSepPunct{\mcitedefaultmidpunct}
{}{\mcitedefaultseppunct}\relax
\EndOfBibitem
\bibitem[Cao \latin{et~al.}(2019)Cao, Zhou, Li, Jia, Liu, Wang, Zhang, Zhang,
  Chen, Xu, and Knutson]{Knutson2019}
Cao,~S.; Zhou,~Z.; Li,~H.; Jia,~M.; Liu,~Y.; Wang,~M.; Zhang,~M.; Zhang,~S.;
  Chen,~J.; Xu,~J.; Knutson,~J.~R. {A fraction of NADH in solution is
  “dark”: Implications for metabolic sensing via fluorescence lifetime}.
  \emph{Chem. Phys. Lett.} \textbf{2019}, \emph{726}, 18–--21\relax
\mciteBstWouldAddEndPuncttrue
\mciteSetBstMidEndSepPunct{\mcitedefaultmidpunct}
{\mcitedefaultendpunct}{\mcitedefaultseppunct}\relax
\EndOfBibitem
\bibitem[Cao \latin{et~al.}(2020)Cao, Li, Liu, Zhang, Wang, Zhou, Chen, Zhang,
  Xu, and Knutson]{Knutson2020}
Cao,~S.; Li,~H.; Liu,~Y.; Zhang,~M.; Wang,~M.; Zhou,~Z.; Chen,~J.; Zhang,~S.;
  Xu,~J.; Knutson,~J.~R. {Femtosecond Fluorescence Spectra of NADH in Solution:
  Ultrafast Solvation Dynamics}. \emph{J. Phys. Chem. B} \textbf{2020},
  \emph{124}, 771--776\relax
\mciteBstWouldAddEndPuncttrue
\mciteSetBstMidEndSepPunct{\mcitedefaultmidpunct}
{\mcitedefaultendpunct}{\mcitedefaultseppunct}\relax
\EndOfBibitem
\bibitem[Gafni and Brand(1976)Gafni, and Brand]{Gafni1976}
Gafni,~A.; Brand,~L. {Fluorescence Decay Studies of Reduced Nicotinamide
  Adenine Dinucleotide in Solution and Bound to Liver Alcohol Dehydrogenase}.
  \emph{Biochemistry} \textbf{1976}, \emph{15}, 3165--3171\relax
\mciteBstWouldAddEndPuncttrue
\mciteSetBstMidEndSepPunct{\mcitedefaultmidpunct}
{\mcitedefaultendpunct}{\mcitedefaultseppunct}\relax
\EndOfBibitem
\bibitem[Krishnamoorthy \latin{et~al.}(1987)Krishnamoorthy, Periasamy, and
  Venkataraman]{Krishnamoorthy1987}
Krishnamoorthy,~G.; Periasamy,~N.; Venkataraman,~B. {On the origin of
  heterogeneity of fluorescence decay kinetics of reduced nicotinamide adenine
  dinucleotide}. \emph{Biochem. Biophys. Research Comm.} \textbf{1987},
  \emph{144}, 387--392\relax
\mciteBstWouldAddEndPuncttrue
\mciteSetBstMidEndSepPunct{\mcitedefaultmidpunct}
{\mcitedefaultendpunct}{\mcitedefaultseppunct}\relax
\EndOfBibitem
\bibitem[Kierdaszuk \latin{et~al.}(1996)Kierdaszuk, Malak, Gryczynski, Callis,
  and Lakowicz]{Kierdaszuk1996}
Kierdaszuk,~B.; Malak,~H.; Gryczynski,~I.; Callis,~P.; Lakowicz,~J.~R.
  {Fluorescence of reduced nicotinamides using one- and two-photon excitation}.
  \emph{Biophys. Chem.} \textbf{1996}, \emph{62}, 1--13\relax
\mciteBstWouldAddEndPuncttrue
\mciteSetBstMidEndSepPunct{\mcitedefaultmidpunct}
{\mcitedefaultendpunct}{\mcitedefaultseppunct}\relax
\EndOfBibitem
\bibitem[Blacker \latin{et~al.}(2019)Blacker, Nicolaou, Duchen, and
  Bain]{Blacker2019}
Blacker,~T.~S.; Nicolaou,~N.; Duchen,~M.~R.; Bain,~A.~J. {Polarized Two-Photon
  Absorption and Heterogeneous Fluorescence Dynamics in NAD(P)H}. \emph{J.
  Phys. Chem. B} \textbf{2019}, \emph{123}, 4705--4717\relax
\mciteBstWouldAddEndPuncttrue
\mciteSetBstMidEndSepPunct{\mcitedefaultmidpunct}
{\mcitedefaultendpunct}{\mcitedefaultseppunct}\relax
\EndOfBibitem
\bibitem[Wu \latin{et~al.}(1995)Wu, Lai, and Houk]{Wu1995}
Wu,~Y.-D.; Lai,~D. K.~W.; Houk,~K.~N. {Transition Structures of Hydride
  Transfer Reactions of Protonated Pyridinium Ion with 1,4-Dihydropyridine and
  Protonated Nicotinamide with 1,4-Dihydronicotinamide}. \emph{J. Am. Chem.
  Soc.} \textbf{1995}, \emph{117}, 4100--4108\relax
\mciteBstWouldAddEndPuncttrue
\mciteSetBstMidEndSepPunct{\mcitedefaultmidpunct}
{\mcitedefaultendpunct}{\mcitedefaultseppunct}\relax
\EndOfBibitem
\bibitem[Hurley and Hammes-Schiffer(1997)Hurley, and
  Hammes-Schiffer]{Hurley1997}
Hurley,~M.~M.; Hammes-Schiffer,~S. {Development of a Potential Surface for
  Simulation of Proton and Hydride Transfer Reactions in Solution: Application
  to NADH Hydride Transfer}. \emph{J. Phys. Chem. A} \textbf{1997}, \emph{101},
  3977--3989\relax
\mciteBstWouldAddEndPuncttrue
\mciteSetBstMidEndSepPunct{\mcitedefaultmidpunct}
{\mcitedefaultendpunct}{\mcitedefaultseppunct}\relax
\EndOfBibitem
\bibitem[Lakowicz(1997)]{Lakowicz97a}
Lakowicz,~J.~R. \emph{Topics in Fluorescence Spectroscopy}; Plenum Press: New
  York, 1997; Vol.~5\relax
\mciteBstWouldAddEndPuncttrue
\mciteSetBstMidEndSepPunct{\mcitedefaultmidpunct}
{\mcitedefaultendpunct}{\mcitedefaultseppunct}\relax
\EndOfBibitem
\bibitem[Herbrich \latin{et~al.}(2015)Herbrich, Al-Hadhuri, Gericke, Shternin,
  Smolin, and Vasyutinskii]{Herbrich15}
Herbrich,~S.; Al-Hadhuri,~T.; Gericke,~K.-H.; Shternin,~P.~S.; Smolin,~A.~G.;
  Vasyutinskii,~O.~S. Two-color two--photon excited fluorescence of indole:
  Determination of wavelength--dependent molecular parameters. \emph{J. Chem.
  Phys.} \textbf{2015}, \emph{142}, 024310\relax
\mciteBstWouldAddEndPuncttrue
\mciteSetBstMidEndSepPunct{\mcitedefaultmidpunct}
{\mcitedefaultendpunct}{\mcitedefaultseppunct}\relax
\EndOfBibitem
\bibitem[Sasin \latin{et~al.}(2018)Sasin, Smolin, Gericke, Tokunaga, and
  Vasyutinskii]{Sasin18}
Sasin,~M.~E.; Smolin,~A.~G.; Gericke,~K.-H.; Tokunaga,~E.; Vasyutinskii,~O.~S.
  Fluorescence anisotropy in indole under two--photon excitation in the
  spectral range 385-–510 nm. \emph{Phys. Chem. Chem. Phys.} \textbf{2018},
  \emph{20}, 19922--19931\relax
\mciteBstWouldAddEndPuncttrue
\mciteSetBstMidEndSepPunct{\mcitedefaultmidpunct}
{\mcitedefaultendpunct}{\mcitedefaultseppunct}\relax
\EndOfBibitem
\bibitem[Sasin \latin{et~al.}(2019)Sasin, Gorbunova, Bezverkhnii, Beltukov,
  Vasyutinskii, and Rubayo-Soneira]{Sasin19}
Sasin,~M.~E.; Gorbunova,~I.~A.; Bezverkhnii,~N.~O.; Beltukov,~Y.~M.;
  Vasyutinskii,~O.~S.; Rubayo-Soneira,~J. Polarized Fluorescence in {NADH}
  Two--Photon Excited by Femtosecond Laser Pulses in the Wavelength Range of
  720-–780 nm. \emph{Tech. Phys. Lett.} \textbf{2019}, \emph{45},
  672--674\relax
\mciteBstWouldAddEndPuncttrue
\mciteSetBstMidEndSepPunct{\mcitedefaultmidpunct}
{\mcitedefaultendpunct}{\mcitedefaultseppunct}\relax
\EndOfBibitem
\bibitem[Shternin \latin{et~al.}(2010)Shternin, Gericke, and
  Vasyutinskii]{Shternin10}
Shternin,~P.~S.; Gericke,~K.-H.; Vasyutinskii,~O.~S. The polarisation of
  two-photon excited fluorescence in rotating Molecules. \emph{Mol. Phys.}
  \textbf{2010}, \emph{108}, 813--825\relax
\mciteBstWouldAddEndPuncttrue
\mciteSetBstMidEndSepPunct{\mcitedefaultmidpunct}
{\mcitedefaultendpunct}{\mcitedefaultseppunct}\relax
\EndOfBibitem
\bibitem[Denicke \latin{et~al.}(2010)Denicke, Gericke, Smolin, Shternin, and
  Vasyutinskii]{Denicke10}
Denicke,~S.; Gericke,~K.-H.; Smolin,~A.~G.; Shternin,~P.~S.;
  Vasyutinskii,~O.~S. Dynamics of Two--Color Two--Photon Excited Fluorescence
  of p--{T}erphenyl: Determination and Analysis of the Molecular Parameters.
  \emph{J. Phys. Chem. A} \textbf{2010}, \emph{114}, 9681--9692\relax
\mciteBstWouldAddEndPuncttrue
\mciteSetBstMidEndSepPunct{\mcitedefaultmidpunct}
{\mcitedefaultendpunct}{\mcitedefaultseppunct}\relax
\EndOfBibitem
\bibitem[McClain(1973)]{McClain73}
McClain,~W. \emph{J. Chem. Phys.} \textbf{1973}, \emph{58}, 324\relax
\mciteBstWouldAddEndPuncttrue
\mciteSetBstMidEndSepPunct{\mcitedefaultmidpunct}
{\mcitedefaultendpunct}{\mcitedefaultseppunct}\relax
\EndOfBibitem
\bibitem[Wan and Johnson(1994)Wan, and Johnson]{Wan94}
Wan,~C.; Johnson,~C. \emph{J. Chem. Phys.} \textbf{1994}, \emph{101},
  10283\relax
\mciteBstWouldAddEndPuncttrue
\mciteSetBstMidEndSepPunct{\mcitedefaultmidpunct}
{\mcitedefaultendpunct}{\mcitedefaultseppunct}\relax
\EndOfBibitem
\bibitem[Zare(1988)]{Zare88b}
Zare,~R.~N. \emph{Angular Momentum}; Wiley: New York, 1988\relax
\mciteBstWouldAddEndPuncttrue
\mciteSetBstMidEndSepPunct{\mcitedefaultmidpunct}
{\mcitedefaultendpunct}{\mcitedefaultseppunct}\relax
\EndOfBibitem
\bibitem[Callis(1993)]{Callis.1993}
Callis,~P.~R. On the theory of two-photon induced fluorescence anisotropy with
  application to indoles. \emph{J. Chem. Phys.} \textbf{1993}, \emph{99},
  27--37\relax
\mciteBstWouldAddEndPuncttrue
\mciteSetBstMidEndSepPunct{\mcitedefaultmidpunct}
{\mcitedefaultendpunct}{\mcitedefaultseppunct}\relax
\EndOfBibitem
\bibitem[Vasyutinskii \latin{et~al.}(2017)Vasyutinskii, Smolin, Oswald, and
  Gericke]{Vasyutinskii2017}
Vasyutinskii,~O.~S.; Smolin,~A.~G.; Oswald,~C.; Gericke,~K.-H. {Polarized
  fluorescence in NADH under two-photon excitation with femtosecond laser
  pulses}. \emph{Opt. Spectr.} \textbf{2017}, \emph{122}, 602--606\relax
\mciteBstWouldAddEndPuncttrue
\mciteSetBstMidEndSepPunct{\mcitedefaultmidpunct}
{\mcitedefaultendpunct}{\mcitedefaultseppunct}\relax
\EndOfBibitem
\bibitem[Thompson \latin{et~al.}(2006)Thompson, Kaiser, and
  Jorgenson]{Thompson2006}
Thompson,~J.~W.; Kaiser,~T.~J.; Jorgenson,~J.~W. Viscosity measurements of
  methanol–water and acetonitrile–water mixtures at pressures up to 3500
  bar using a novel capillary time-of-flight viscometer. \emph{J. Chromatogr.
  A} \textbf{2006}, \emph{1134}, 201--209\relax
\mciteBstWouldAddEndPuncttrue
\mciteSetBstMidEndSepPunct{\mcitedefaultmidpunct}
{\mcitedefaultendpunct}{\mcitedefaultseppunct}\relax
\EndOfBibitem
\bibitem[Wu and Houk(1993)Wu, and Houk]{Wu1993}
Wu,~Y.~D.; Houk,~K.~N. {Theoretical Study of Conformational Features of NAD+
  and NADH Analogs: Protonated Nicotinamide and 1,4-Dihydronicotinamide}.
  \emph{J. Org. Chem.} \textbf{1993}, \emph{58}, 2043--2045\relax
\mciteBstWouldAddEndPuncttrue
\mciteSetBstMidEndSepPunct{\mcitedefaultmidpunct}
{\mcitedefaultendpunct}{\mcitedefaultseppunct}\relax
\EndOfBibitem
\bibitem[Kumar \latin{et~al.}(2010)Kumar, Jaiswal, Singh, Srivastav, Singh,
  Yadav, and Yadav]{Kumar2010}
Kumar,~M.; Jaiswal,~S.; Singh,~R.; Srivastav,~G.; Singh,~P.; Yadav,~T.;
  Yadav,~R. {Ab initio studies of molecular structures, conformers and
  vibrational spectra of heterocyclic organics: I. Nicotinamide and its
  N-oxide}. \emph{Spectrochim. Acta A} \textbf{2010}, \emph{75}, 281--292\relax
\mciteBstWouldAddEndPuncttrue
\mciteSetBstMidEndSepPunct{\mcitedefaultmidpunct}
{\mcitedefaultendpunct}{\mcitedefaultseppunct}\relax
\EndOfBibitem
\bibitem[Frisch and et~al.()Frisch, and et~al.]{GAUSSIAN09}
Frisch,~M.~J.; et~al., Gaussian 09, \uppercase{R}evision \uppercase{D}.01.
  \uppercase{G}aussian, Inc., Wallingford, CT, 2013\relax
\mciteBstWouldAddEndPuncttrue
\mciteSetBstMidEndSepPunct{\mcitedefaultmidpunct}
{\mcitedefaultendpunct}{\mcitedefaultseppunct}\relax
\EndOfBibitem
\bibitem[Kovacs \latin{et~al.}(2017)Kovacs, Dobrowolski, Ostrowski, and
  Rode]{Kovacs2017}
Kovacs,~A.; Dobrowolski,~J.~C.; Ostrowski,~S.; Rode,~J.~E. {Benchmarking
  density functionals in conjunction with Grimme's dispersion correction for
  noble gas dimers (Ne2, Ar2, Kr2, Xe2, Rn2)}. \emph{Int. J. Quantum Chem.}
  \textbf{2017}, \emph{117}, 1--13\relax
\mciteBstWouldAddEndPuncttrue
\mciteSetBstMidEndSepPunct{\mcitedefaultmidpunct}
{\mcitedefaultendpunct}{\mcitedefaultseppunct}\relax
\EndOfBibitem
\bibitem[Smith and Tanner(2000)Smith, and Tanner]{Smith2000}
Smith,~P.~E.; Tanner,~J.~J. {Conformations of nicotinamide adenine dinucleotide
  (NAD+) in various environments}. \emph{J. Molec. Recog.} \textbf{2000},
  \emph{13}, 27--34\relax
\mciteBstWouldAddEndPuncttrue
\mciteSetBstMidEndSepPunct{\mcitedefaultmidpunct}
{\mcitedefaultendpunct}{\mcitedefaultseppunct}\relax
\EndOfBibitem
\bibitem[Scott \latin{et~al.}(1970)Scott, Spencer, Leonard, and
  Weber]{Scott1970}
Scott,~T.~G.; Spencer,~R.~D.; Leonard,~N.~J.; Weber,~G. {Emission Properties of
  NADH. Studies of Fluorescence Lifetimes and Quantum Efficiencies of NADH,
  AcPyADH, and Simplified Synthetic Models}. \emph{J. Am. Chem. Soc.}
  \textbf{1970}, \emph{92}, 687--695\relax
\mciteBstWouldAddEndPuncttrue
\mciteSetBstMidEndSepPunct{\mcitedefaultmidpunct}
{\mcitedefaultendpunct}{\mcitedefaultseppunct}\relax
\EndOfBibitem
\bibitem[Nakabayashi \latin{et~al.}(2014)Nakabayashi, Islam, Li, Yasuda, and
  Ohta]{Nakabayashi2014}
Nakabayashi,~T.; Islam,~M.~S.; Li,~L.; Yasuda,~M.; Ohta,~N. {Studies on
  external electric field effects on absorption and fluorescence spectra of
  NADH}. \emph{Chem. Phys. Lett.} \textbf{2014}, \emph{595-596}, 25--30\relax
\mciteBstWouldAddEndPuncttrue
\mciteSetBstMidEndSepPunct{\mcitedefaultmidpunct}
{\mcitedefaultendpunct}{\mcitedefaultseppunct}\relax
\EndOfBibitem
\bibitem[Dote \latin{et~al.}(1981)Dote, Kivelson, and Schwartz]{Dote1981}
Dote,~J.~L.; Kivelson,~D.; Schwartz,~R.~N. A Molecular Quasi-Hydrodynamic
  Free-Space Model for Molecular Rotational Relaxation In Liquids. \emph{J.
  Phys. Chem.} \textbf{1981}, \emph{85}, 2169--2180\relax
\mciteBstWouldAddEndPuncttrue
\mciteSetBstMidEndSepPunct{\mcitedefaultmidpunct}
{\mcitedefaultendpunct}{\mcitedefaultseppunct}\relax
\EndOfBibitem
\bibitem[Anderton and Kauffman(1994)Anderton, and Kauffman]{Anderton1994}
Anderton,~R.~M.; Kauffman,~J.~F. Temperature-Dependent Rotational Relaxation of
  Diphenylbutadiene in n-Alcohols: A Test of the Quasihydrodynamic Free Space
  Model. \emph{J. Phys. Chem.} \textbf{1994}, \emph{98}, 12117--12124\relax
\mciteBstWouldAddEndPuncttrue
\mciteSetBstMidEndSepPunct{\mcitedefaultmidpunct}
{\mcitedefaultendpunct}{\mcitedefaultseppunct}\relax
\EndOfBibitem
\bibitem[Hu and Zwanzig(1974)Hu, and Zwanzig]{Hu1974}
Hu,~C.-M.; Zwanzig,~R. Rotational friction coefficients for spheroids with the
  slipping boundary condition. \emph{J. Chem. Phys.} \textbf{1974}, \emph{60},
  4354--4357\relax
\mciteBstWouldAddEndPuncttrue
\mciteSetBstMidEndSepPunct{\mcitedefaultmidpunct}
{\mcitedefaultendpunct}{\mcitedefaultseppunct}\relax
\EndOfBibitem
\bibitem[Edward(1970)]{Edward1970}
Edward,~J.~T. Molecular volumes and the Stokes-Einstein equation. \emph{J.
  Chem. Ed.} \textbf{1970}, \emph{47}, 261--270\relax
\mciteBstWouldAddEndPuncttrue
\mciteSetBstMidEndSepPunct{\mcitedefaultmidpunct}
{\mcitedefaultendpunct}{\mcitedefaultseppunct}\relax
\EndOfBibitem
\bibitem[Qin \latin{et~al.}(2009)Qin, Huang, and Suganthan]{Qin2009}
Qin,~A.~K.; Huang,~V.~L.; Suganthan,~P.~N. {Differential Evolution Algorithm
  With Strategy Adaptation for Global Numerical Optimization}. \emph{IEEE T.
  Evolut. Comput.} \textbf{2009}, \emph{13}, 398--417\relax
\mciteBstWouldAddEndPuncttrue
\mciteSetBstMidEndSepPunct{\mcitedefaultmidpunct}
{\mcitedefaultendpunct}{\mcitedefaultseppunct}\relax
\EndOfBibitem
\end{mcitethebibliography}

\end{document}